\documentclass[aoas-arxiv]{imsart}

\RequirePackage{amsthm,amsmath,amsfonts,amssymb}
\RequirePackage[authoryear]{natbib}
\RequirePackage[colorlinks,citecolor=blue,urlcolor=blue]{hyperref}
\RequirePackage{graphicx}
\usepackage{xr}
\usepackage{lscape, pdflscape}
\usepackage{hyperref, enumitem}
\usepackage{mathtools, booktabs}
\usepackage{csquotes}
\usepackage{tikz, tikz-network}

\startlocaldefs

\input{math_commands.tex}

\newtheorem{theorem}{Theorem}

\newtheorem{definition}[theorem]{Definition}

\newtheorem{notation}[theorem]{Notation}

\newtheorem{remark}{Remark}

\newcommand{\N}{\ensuremath{\mathbb{N}}}

\newcommand{\GPDText}{\ensuremath{\mathrm{GPD}}}
\newcommand{\GPD}{\GPDText}
\newcommand{\MGPD}{\ensuremath{\mathrm{MGPD}}}
\newcommand{\GPDTextwithspace}{\ensuremath{\mathrm{GPD \ }}}

\newcommand{\sumtext}{\ensuremath{\mathrm{sum}}}
\newcommand{\factual}{\ensuremath{\mathrm{f}}}
\newcommand{\counterfactual}{\ensuremath{\mathrm{cf}}}

\newcommand{\indicator}{\ensuremath{\mathbb{I}}}

\newcommand{\UU}{\ensuremath{\mathbb{U}}} 
\newcommand{\YY}{\ensuremath{\mathbb{Y}}} 
\newcommand{\XX}{\ensuremath{\mathbb{X}}} 
\newcommand{\ZZ}{\ensuremath{\mathbb{Z}}} 

\newcommand{\proba}{\ensuremath{{\text{P}}}} 

\newcommand{\UEV}{\ensuremath{\mathrm{UEV}}} 

\newcommand{\PA}{\ensuremath{\mathrm{PA}}} 

\newcommand{\indep}{\perp \!\!\! \perp}

\newcommand{\PN}{\ensuremath{ \mathrm{PN}}}
\newcommand{\PS}{\ensuremath{ \mathrm{PS}}}
\newcommand{\PNS}{\ensuremath{ \mathrm{PNS}}}
\newcommand{\PC}{\ensuremath{ \mathrm{PC}}}
\newcommand{\mBICV}{\ensuremath{ \mathrm{mBICV}}}
\newcommand{\ACE}{\ensuremath{ \mathrm{ACE}}}

\endlocaldefs

\begin{document}

\begin{frontmatter}
\title{Extreme event propagation using counterfactual theory and vine copulas}
\runtitle{Extreme event propagation}

\begin{aug}
\author[A]{\fnms{Valentin} \snm{Courgeau}\ead[label=e1]{valentin.courgeau15@imperial.ac.uk}},
\and
\author[B]{\fnms{Almut E.D.} \snm{Veraart}\ead[label=e2]{a.veraart@imperial.ac.uk}}
\address[A]{180 Queen's Gate, London, SW7 2AZ, UK, \printead{e1}}
\address[B]{180 Queen's Gate, London, SW7 2AZ, UK, \printead{e2}}
\end{aug}

\begin{abstract}
 Understanding multivariate extreme events play a crucial role in managing the risks of complex systems since extremes are governed by their own mechanisms. Conditional on a given variable exceeding a high threshold (e.g.\ traffic intensity), knowing which high-impact quantities (e.g\ air pollutant levels) are the most likely to be extreme in the future is key. This article investigates the contribution of marginal extreme events on future extreme events of related quantities. We propose an Extreme Event Propagation framework to maximise counterfactual causation probabilities between a known cause and future high-impact quantities. Extreme value theory provides a tool for modelling upper tails whilst vine copulas are a flexible device for capturing a large variety of joint extremal behaviours. We optimise for the probabilities of causation and apply our framework to a London road traffic and air pollutants dataset. We replicate documented atmospheric mechanisms beyond linear relationships. This provides a new tool for quantifying the propagation of extremes in a large variety of applications.
\end{abstract}


\end{frontmatter}


\section{Introduction}
Quantifying dependencies between extremes is essential in analysing risk scenarios and extreme chain reactions in applications, e.g.\ extreme air pollution, meteorology, hydrology \citep{dutfoy2014multivariate, desario2013climatechangeExtremes, bevacqua2017multivariate} or financial risk management \citep{embrechts2013modelling}. 
Causal relationships in the context of time series are well-established \citep{eichler2007causal, eichler2013causalMultipleTS}, most notably with the Granger causality \citep{granger1980causality, granger1988causality} while causality for extremes has shown promising results recently \citep{kiriliouk2020ClimateExtremeEventAttribution, hannart2016causal, hannart2018probabilities}. 
However, in this case, the temporal evolution of extremes is usually modelled backwards: for instance, the celebrated Extreme Event Attribution (EEA) methodology \citep{allen2003liabilityClimateChange} attributes a particular extreme event to its potential causes \citep{angelil2017independent, philip2020protocolEEA, trenberth2015AttributionClimateExtremes}, especially for climate-related applications. 
In this article, we provide a forward-looking scheme for extremes to investigate the causal impact of an extreme event on a high-impact event through time which we call the \emph{Extreme Event Propagation (EEP)} framework. This builds upon the peaks-over-thresholds extreme value theory \citep{nerc1975, davison1990models, pickands1971two} and the counterfactual causal theory \citep{pearl1999probaOfCausation}. 

The peaks-over-thresholds literature takes root in the idea that high values have their own mechanisms and should be treated separately from ordinary values as only those large values provide insights about other extreme values \citep{rootzen2018MultivariatePoTModels}. To differentiate ordinary values from extreme ones, a threshold is defined and values above it are considered to be \emph{peaks}. The excess between the threshold and the peaks have been shown to converge to a generalised Pareto distribution (GPD) \citep{dehaan2977limitTheoryExtremes, balkema1974residual, pickands1971two} as the threshold approaches the distribution's endpoint. A multivariate extension of the GPD  (MGPD) was introduced in \cite{beirlant2004statistics} \& \cite{rootzen2006mgpd}. We note that spatial extremes have been instrumental in the development of those extensions \citep{wadsworth2018spatial,  bacro2020spacetime}, supported by a family of models with an ever-growing distributional and asymptotic flexibility \citep{wadsworth2017ModellingAcrossDependenceClasses, rootzen2018MultivariatePoTModels, kiriliouk2019PoTMultivariateGPD}. We rely on a copula-based MGPD definition presented in \cite{falk2019GeneralizedParetoCopulas}, where marginal distributions have GPD upper tails and are all together linked through a threshold-stable copula structure called a generalised Pareto copula. 

 On the other hand, the counterfactual causal theory \citep{pearl1999probaOfCausation} relies on a cause event and an impact event and opposes two versions of the world: the factual world where the cause happened, and the counterfactual one where it did not. We compare those settings through three combined probabilities, called \emph{probabilities of causation}, to quantify and potentially maximise the necessary or sufficient nature of the cause on the impact event, see \cite{naveau2020statistical} for a recent statistical review. \cite{hannart2018probabilities} and \cite{kiriliouk2020ClimateExtremeEventAttribution} apply this approach to extreme values for atmospheric applications. The EEP framework is an extension of their idea where we model both the cross-sectional and temporal dependencies, bringing the analysis a step closer to handling realistic risk management situations. Alternatively, \cite{gnecco2019causal} define a causal tail coefficient that captures asymmetries in the extremal dependence of two random variables and \cite{mhalla2020causalmechanism} construct an information-theoretic statistic to uncover causal links.

Although the cause and impact events can be tailored to the application at hand \citep{philip2020protocolEEA, hannart2018probabilities}, we also describe our framework with parametrised cause and impact events. For instance, \cite{bevacqua2017multivariate} introduces compound events where the individual variables may not be extreme themselves but their joint occurrence causes an extreme impact---a case that is therefore covered in the EEP framework. On the other hand, \cite{hannart2016causal} maximises a probability of causation with respect to an extreme temperature threshold, delimiting high values from ordinary ones. This also allows for the discovery of causal links between the cause and individual variables through time as an alternative to the pairwise approaches \citep{peters2014causaldiscovery, mhalla2020causalmechanism, gnecco2019causal} or sparse structures for extremes \citep{engelke2021reviewsparse, engelke2020structure, engelke2020graphical} that extract the most significant extremal pairwise links in a tractable manner.

Modelling extremal behaviour accurately requires flexible dependence structures as formalised by the asymptotic dependence \citep{wadsworth2017ModellingAcrossDependenceClasses, huser2019UnknownDependenceClass}, regularly varying distributions \citep{mikosh2006copulas, weng2012characterisationMRV} or sparse structures for high-dimensions \citep{engelke2021reviewsparse, engelke2020graphical}. Although the EEP framework remains model-agnostic in its current formulation, vine copulas \citep{bedford2002vines}---a family of hierarchical pairwise graphical copulas---are a balance between adjustable extremal properties \citep{joe2010tail}, the ability to capture non-linear relationships \citep{graler2014modelling, erhardt2015r} and scalability capabilities \citep{nagler2018model, nagler2020stationary}. It also offers a conditional sampling mechanism \citep[App.\ B]{bevacqua2017multivariate}, a key ingredient for counterfactual reasoning \citep[Section 2.b]{eichler2013causalMultipleTS}.

After a short presentation of counterfactual theory and time-series causality, we introduce the components of the EEP framework in Section \ref{section:eep}: the cause and impact events as well as the counterfactual probabilistic setting to understand their causal link. In Section \ref{section:multivariate-extremes}, we recall the properties of interest for multivariate extremes modelling related to copulas and define the semiparametric and counterfactual marginal model used. The model inference is deferred to Section \ref{section:inference} where marginal parameters are obtained by maximum likelihood whilst extreme thresholds are chosen by sequential hypothesis testing. We also mention different inference approaches for vine copulas such as a tree structure selection algorithm and model selection criteria. Tail probabilities are essential quantities in counterfactual settings and we introducing a marginal transformation approach to handle the cross-sectional comparisons of potentially different quantities in nature.  Finally, Section \ref{section:london-air-pollution-dataset} is devoted to presenting a case study about the impact of high road traffic on air pollutant concentration peaks in the following six hours on Bloomsbury Road, London (UK). An overview of causality for time series and shortlisted definitions from the counterfactual causal theory and vine copulas literature are presented in Appendix \ref{section:time-series-causality}-- \ref{section:archimedean-regular-vine-copulas}.

\section{Extreme Event Propagation}
\label{section:eep}
Starting with \cite{allen2003liabilityClimateChange}, the EEA literature focuses on determining if \emph{climate change influenced the frequency, likelihood, and/or severity of individual extreme events} \citep[p.\ 525]{swain2020EEAClimateChange}. 

We are interested in modelling the opposite, that is, how the conditions linked to an extreme event taking place can propagate through time and cross-sections to increase the probability of another, potentially higher-impact, extreme event. Also, we quantify which of the marginals involved in this impact event are more likely to reach extreme levels in the future. We call this the \emph{Extreme Event Propagation (EEP)} framework.

\subsection{Foundations}
\label{section:foundations}
EEP applies the causal counterfactual theory \citep{pearl1999probaOfCausation} to assess how a quantity above a high threshold at a given time can impact a collection of target variables to become extreme later in time. 
 
 In this article, we oppose a factual version of the world (i.e.\ where the cause intervenes on the system under study) and counterfactual version (i.e.\ where it does not intervene) version to observe and potentially maximise a given probability of causation of the potential cause on a parametrised impact event. We decompose the EEP framework into three steps: 
  \begin{enumerate}[label=(\roman*)]
  	\item defining the cause i.e.\ the \emph{factual} and \emph{counterfactual} worlds;
  	\item defining the (parametrised) \emph{impact} event; 
  	\item comparing those settings through a probabilistic causal framework; 
  \end{enumerate}
    analogously to the EEA framework \citep{swain2020EEAClimateChange, philip2020protocolEEA}.

\subsection{Notations}
Let $I = \{1,\dots,d\}$ be an index set for the $d \in \N$ variables. On a filtered probability space $(\Omega, \mathcal{F}, \{\mathcal{F}_t,\ t \in \sN\}, \proba)$, we consider an adapted and strictly stationary $d$-dimensional process 
$$\XX_t=(\ermX^{1}_t, \dots, \ermX^{d}_t)^\top,\quad t \in \N,$$ where we simplify the usual marginal notation $\ermX^{(i)}$ into $\ermX^{i}$ for conciseness. We write $F$ the cdf (resp.\ $f$ the pdf) of $\XX$; $F^{i}$ the cdf (resp.\ $f^{i}$ the pdf) of $\ermX^{i}$ for $i \in I$. In addition, we write the quantile function (or generalised inverse) of a univariate distribution $F^{i}$ by $Q^i(u) \overset{\Delta}{=} \inf\left\{t\in \R : F(t) \geq u\right\}$ for $u \in [0,1]$. For an offset index $k\in\N$, we denote the forward-looking $k$-stack of $\XX_{t}, \dots, \XX_{t+k}$ by 
 $$\XX_t(k) \overset{\Delta}{=} (\XX_{t}^\top,\dots,\XX_{t+k}^\top)^\top \in \R^{(k+1)d}.$$
The integral transform is given by $\ermU^{i}_t \overset{\Delta}{=} F^{i}(\ermU^{i}_t) \sim \mathcal{U}(0,1),$
and we define
$$\UU_t \overset{\Delta}{=} (U^i_t,\ i \in I) \in (0,1)^d, \quad \text{and} \quad  \UU_t(k) \overset{\Delta}{=} (\UU_{t}^\top, \dots, \UU_{t+k}^\top)^\top \in (0,1)^{(k+1)d}.$$
The generalised Pareto distribution (GPD) with threshold $u$, shape $\gamma \in \R$ and scale $\sigma > 0$, denoted $\GPD(u,\gamma,\sigma)$, has a pdf given by 
$$f(x;\gamma,\sigma,u)=\sigma^{-1}\left(1+\frac{\gamma}{\sigma}(x-u)\right)_+^{-1/\gamma},$$
where $x_+ \overset{\Delta}{=} \max(x,0)$. If $u=0$, we simply denote this distribution by $\GPD(\gamma,\sigma)$.

\subsection{The cause}
\label{section:cause}
We are interested in quantifying the contribution of individual variables on the impact event and consider the $i$-th \emph{marginal extreme event} denoted by 
\begin{equation}
	\label{eq:marginal-extreme-event}
	\mC_t^{i} \overset{\Delta}{=} \mC_t(\mu^{i}, \boldsymbol{e}^i_d) = \{X^{i}_t > \mu^{i}\}, \quad \text{for some $\mu^{i} \in \R$,} \quad i \in I,
\end{equation}
where $\boldsymbol{e}^i_d = (0,\dots,0,1,0,\dots,0)^\top \in \R^d$ is the $i$-th standard unit vector in $\R^d$. The specific value chosen for $\mu^{i}$ is a threshold value above which $\ermX^{i}$ is said to be \emph{extreme} and is formally defined in Section \ref{section:mgpd}.

\subsection{The impact event}
\label{section:impact-event}
We define an impact function $h = h(\boldsymbol{x}; \vw):\R^{p+m} \rightarrow \R$ for $p,m \in \N$ where $\boldsymbol{x} \in \R^p$ are the variables whilst the vector $\vw\in\R^m$ parametrise the function $h$ \citep{bevacqua2017multivariate}. To simplify the notation, we allow for different values of $p$ and $m$ in the definition of $h$ when the context is clearly defined.

We consider the event of $h\left(\XX_{t}(k);\vw\right)$ taking its value above a high threshold $v \in \sR$ for some parameter vector $\vw\in (0,\infty)^m$, denoted
$$\mE_t(v, \vw, k; h) \overset{\Delta}{=} \left\{ h(\XX_{t+1}(k);\vw) > v\right\}, \quad \vw \in [0,\infty)^m, \quad k\in\sN,$$
which we call the \emph{impact} event of $\XX$ at time $t+k$ with respect to the threshold $v$ and impact function $h$.

\begin{remark}
	The impact function was originally introduced by \cite{bevacqua2017multivariate} to study compound events where individual variables may not be extreme but a collection of those variables can cause some extreme impact. 
\end{remark}

\begin{definition}
	 A measurable function $L:[0,\infty) \mapsto [0,\infty)$ for which the limit $ g(a)=\lim_{x\rightarrow\infty}L(ax)/L(x)$ exists and is finite for all $a > 0$, is called regularly varying. If $g(a)=1$ for all $a>0$, then $L$ is called slowly varying.
\end{definition}

\subsubsection{Linear impact events}
The multivariate regular variation (MRV) property is an important tool to model multivariate heavy-tailed phenomena \citep{dehaan2977limitTheoryExtremes}. See Chapter 5, \cite{resnick1987extremes} for additional information on the MRV property. We leverage a key characterisation  of the MRV property \citep{basrak2002characMRV} which states that any random vector $\YY\in \R^m$ satisfies the MRV property if and only if every linear combination of the random vector is (univariate) regularly varying: that is, if there exists a $\beta > 0$ and a slowly varying function $L$ such that, for all $\vw \in \R^m$, the limit
$$\lim_{u\rightarrow\infty}\frac{\proba(\vw^\top \YY > u)}{u^{-\beta}L(u)} \quad \text{exists,}$$
 and there exists some $\vw_0\neq \boldsymbol{0}$ such that the limit is non-zero. Therefore, if the MRV property holds for the multivariate distribution of $\XX_t(k)$, then, $\vw^\top \XX_{t+1}(k)$ is regularly varying as a marginal distribution, irrespective of the weight vector $\vw$. 

Therefore,  we only investigate the weighted sum impact function which is denoted by
$$h_\sumtext(\vy, \vw) = \vw^\top \vy,\quad \text{for }\vy \in \R^m.$$
Conveniently, this helps us separate the methodology that we introduce with the modelling of impact events which is beyond the scope of this article. We define the family of linear impact events with all marginals between times $t+1$ and $t+k$ as follows
\begin{equation}
	\label{eq:linear-impact}
	\mE_t(v, \vw,k) \overset{\Delta}{=} \left\{ \vw^\top \XX_{t+1}(k) > v\right\}, \quad \vw \in [0,\infty)^{kd}, \quad k \in \sN,
\end{equation}
where $\XX_{t+1}(k) = (\XX_{t+1}^\top, \dots, \XX_{t+k}^\top)^\top \in \R^{kd}$ such that $\mE_t(v, \vw,k) \in \gF_{t+k}$.

\subsubsection{Examples}
\label{section:impact-event:examples}
We denote by $\boldsymbol{e}^{i}_m = (0,\dots,0,1,0,\dots,0)^\top \in \R^m$ the $i$-th standard unit vector in $\R^m$ and by $\otimes$ the Kronecker product. 
\paragraph{Cross-sectional impact event} For $l\in\{1,\dots,k\}$ and a weight vector $\vw^l_d \in [0,\infty)^d$, the impact event of the cross-sectional combination of $\XX$ at time $t+l$ is given by
\begin{equation}
\label{eq:cross-section-impact}
	\mE_{t+l}(v, \vw^l_d) \overset{\Delta}{=} \mE_{t}(v, \ve^l_k \otimes \vw^l_d, k) = \left\{{\vw^l_d}^\top \XX_{t+l} > v\right\} = \left\{ \sum_{i=1}^d \vw^{l,i}_d X^{i}_{t+l} > v\right\},
\end{equation}
similarly to the framework of \cite{kiriliouk2020ClimateExtremeEventAttribution} such that $\mE_t(v, \vw^l_d,k) \in \gF_{t+l} \subseteq \gF_{t+k}$.

\paragraph{Timewise marginal impact event} For $j\in\{1,\dots,d\}$ and a weight vector $\vw^j_k \in [0,\infty)^{k}$, the $j$-th timewise marginal impact event between $t+1$ and $t+k$ is defined by
\begin{equation}
\label{eq:time-wise-impact}
	\mE^{j}_t(v, \vw^j_k, k) \overset{\Delta}{=}  \mE_t(v, \vw^j_k \otimes \ve^j_d, k)= \left\{{\vw^j_k}^\top \XX^j_{t+1}(k) > v\right\} = \left\{ \sum_{l=1}^k \vw^{j,l}_k X^{j}_{t+l} > v\right\},
\end{equation}
where $\XX^j_{t+1}(k) = (X^j_{t+1}, \dots, X^j_{t+k})^\top \in \R^k$ and $\mE^{j}_t(v, \vw^j_k, k) \in \gF_{t+k}$. In this case, we are interested in the temporal changes specific to the $j$-th marginal.

\subsection{Causal probabilities}
\label{section:causal-probabilities-eep}
Consider a cause event $\mC_t \in \gF_{t}$ from Section \ref{section:cause} (e.g.\ peak traffic level at time $t$) and an impact event $\mE_{t}(k) \in \gF_{t+k}$ from Section \ref{section:impact-event} (e.g.\ air pollutant levels up to time $t+k$).
\subsubsection{Probabilities of causation} Using the do-notation of \cite{pearl1999probaOfCausation}, the $\PC$s are defined and given as follows:
\begin{align*}
	\PN(k) \ &\overset{\Delta}{=} \ \proba\big(\mE(k)^c\ |\ do(\mC^{c}), \mC, \mE(k)\big)  \quad &&\textit{(necessary),}\\
	\PS(k) \ &\overset{\Delta}{=} \  \proba\big(\mE(k)\ |\ do(\mC), \mC^{c}, \mE(k)^{c}\big), \quad &&\textit{(sufficient),}\\
	\PNS(k) \ &\overset{\Delta}{=} \ \proba\big(\{\mE(k)\ | \ do(\mC)\} \cap \{\mE(k)\ |\ do(\mC^{c})\}\big), \quad &&\textit{(sufficient and necessary).}
\end{align*}
The probability of \emph{necessary} causation $(\PN)$ represents how likely it is that the event $\mE(k)$ has not occurred had the cause $\mC$ not occurred itself, given that both the event $\mE(k)$ and the cause $\mC$ have actually occurred. On the other hand, the probability of \emph{sufficient} causation $(\PS)$ quantifies how probable the impact event $\mE(k)$ would have occurred in the presence of the cause, given that neither the impact event and the cause have occurred. Finally, the probability of \emph{necessary and sufficient} causation $(\PNS)$ represents how likely it is that the impact event occurs in the presence of the cause and would not occur in the absence of the cause.

\subsubsection{Counterfactual probabilities}
We derive the probability of necessary causation with respect to impact events described in Section \ref{section:impact-event}. For the corresponding cause $\mC^{i}$, the probability of $\mE(v,\vw,k)$ is denoted by 
\begin{align*}
	p^{(i)}_\factual(v, \vw; k) \ &\overset{\Delta}{=} \quad \proba\left(\mE(v,\vw,k)\ |\ \mC^{i}\right),    &&\text{in the \emph{factual} world},\\
	p^{(i)}_{\counterfactual}(v, \vw; k) \ &\overset{\Delta}{=} \quad \proba\left(\mE(v,\vw,k)\ |\ \mC^{i,c}\right),  &&\text{in the \emph{counterfactual} world},
\end{align*}
where the time index $t$ is omitted since $\XX$ is stationary. Those are called factual and counterfactual probabilities, respectively: that $\vw^\top\XX_{1}(k) = \sum_{l=1}^{k} \sum_{j=1}^d w_{(l-1)d +j} \XX^j_{l}$ is above a threshold $v$ given that the $i$-th dimension was extreme (resp.\ not extreme) at time $0$. 

 Furthermore, we assume that $\mC^{i}$ is exogenous with respect to $\mE(k)$ and that $\mE(k)$ is monotonous with respect to $\mC^{i}$ \citep[Def.\ 13 \& 14]{pearl1999probaOfCausation}, such that $\PN$, $\PN$ and $\PNS$ are all identifiable \citep[Def.\ 15]{pearl1999probaOfCausation}. For $v \in \R$ and $\vw\in [0,\infty)^{kd}$, the corresponding (counter)factual probabilities are defined by
\begin{align}
\label{eq:linear-factual}
	p^{(i)}_\factual(v, \vw; k) &= \proba\left(\vw^\top \XX_1(k) > v \ \big|\ \ermX^{i}_0 > \mu^{i} \right),\\
\label{eq:linear-counterfactual}
p^{(i)}_\counterfactual(v, \vw; k) & = \proba\left(\vw^\top \XX_1(k) > v\big|\ \ermX^{i}_0 \leq \mu^{i} \right),
\end{align}
that is, the probability that $\vw^\top\XX_{1}(k) = \sum_{l=1}^{k} \sum_{j=1}^d w_{(l-1)d +j} \XX^j_{l}$ is above a threshold $v$ given that the $i$-th dimension was extreme (resp.\ not extreme) at time $0$. 
For a general weight vector $\vw\in [0,\infty)^{kd}$, the corresponding PCs are given by
\begin{align*}
	\PN^{(i)}(v,\vw;k) = \left(1-\frac{p^{(i)}_\counterfactual(v, \vw; k)}{p^{(i)}_\factual(v, \vw; k)} \right)_+,&\quad \quad 
	\PS^{(i)}(v,\vw;k) = \left(1-\frac{1-p^{(i)}_\factual(v, \vw; k)}{1-p^{(i)}_\counterfactual(v, \vw; k)}\right)_+,
\end{align*}
and
$$
\PNS^{(i)}(v,\vw;k) = \Big(p^{(i)}_\counterfactual(v, \vw; k) - p^{(i)}_\factual(v, \vw; k) \Big)_+,
$$
where $x_+ = \max(x,0)$ as given in Eq.\ (8), \cite{hannart2016causal}.
 However, the PCs quantify in three different ways the relationship of the cause on the impact events which we believe are better suited to reflect the complex mechanisms involved in the propagation of extremes. That is, they gain value when presented jointly (see Section \ref{section:london-air-pollution-dataset} and \cite{pearl1999probaOfCausation}). 

\begin{remark}
	\cite{hannart2018probabilities} studies the impact of PNS to uncover causal links between a cause and a parametrisable impact event whilst \cite{kiriliouk2020ClimateExtremeEventAttribution} focus on maximising PN to attribute weather precipitation to anthropogenic forcings (i.e.\ human influence). 
\end{remark}

To estimate factual and counterfactual probabilities, we model the relationships between the marginals at different times using stationary vine copulas \citep{nagler2020stationary}, see Section \ref{section:archimedean-regular-vine-copulas}. In the following section, we detail the motivation behind the use of this model class as well as the different modelling approaches and the estimation schemes available to compute such probabilities.

\section{Modelling multivariate extreme values}
\label{section:multivariate-extremes}
For some fixed $k\in\N$, we consider a strictly stationary $\R^d$-valued time series $(\XX_{t}, \ t \in \N)$ whose $k$-stack $(\XX_{t}(k), \ t \in \N)$ has a distribution function denoted by $F$ in the $\delta$-neighbourhood of a Multivariate Generalised Pareto Distribution (MGPD) \citep[Definition 1]{falk2019GeneralizedParetoCopulas} as recalled in the following section. We then introduce our modelling approach based on Archimedean stationary vine copulas \citep{bedford2001vines, bedford2002vines}.

\subsection{Tail dependence modelling}
\cite{heffernan2004conditional} set a precedent by outlining a flexible multivariate extreme value framework capable to capture a wide range of tail dependencies. We introduce the dependency measures and succinct theoretical justifications that stationary vine copulas are suitable to model those asymptotic behaviours.

\subsubsection{Asymptotic dependence of extremes}
\label{section:asymptotic-dependence-tail-coefficient}
Similarly to an upper tail dependence coefficient, we introduce a modified extremal correlation \citep[Eq.\ (9)]{engelke2020structure} at horizon $h \in \N$ on a set index $S \subseteq I$ as follows
\begin{equation*}
    \chi^{(i)}(S;h) \overset{\Delta}{=} \lim_{u \uparrow 1}\proba\left(X^{j}_{h} >  Q^{j}(u), \ \forall j \in S\ \big| \ X^{i}_{0} > Q^{i}(u)\right), \quad \forall i \in I,
\end{equation*}
and, in particular, the pairwise tail coefficients are given by $\chi^{(i,j)}(h) \overset{\Delta}{=} \chi^{(i)}(\{j\};h)$ for $i,j\in I$. The latter quantifies the joint extremal behaviour of the $j$-th component as the $i$-th component becomes high with a time difference $h$. 

If $\chi^{(i,j)}(h) = 0$, we say that $X^i$ and $X^j$ are asymptotically independent at horizon $h$. Otherwise, if $0 < \chi^{(i,j)}(h) < 1$, they are said to be asymptotically dependent and for $\chi^{(i,j)}(h) = 1$ they are completely extremal dependent \citep{coles1999dependenceMeasures, schlather2003dependenceMeasure}. However, the tail coefficients do not characterise the extremal dependence structure of $\XX_t(k)$, see Section 8.1, \cite{mikosh2006copulas}. \textbf{add details back?}

Some recent models have the limitation that either $\chi^{(i,j)}(h) = 0$ or $\chi^{(i,j)}(h) > 0$ for all $i,j\in I$ and some fixed $h\in\N$ \citep{huser2019UnknownDependenceClass, kiriliouk2019PoTMultivariateGPD} whereas some models avoid this issue \citep{Winter2017, winter2016modelling} by allowing for a mixture of both zero and positive pairwise tail coefficients. Similarly, our model goes in this direction as regular vines adapt their structure to accommodate some pairwise tail coefficients to be zero and others to be positive (Section \ref{section:tail-dependence-function}).

\subsubsection{Tail dependence functions}

Suppose $C=C(\vu)$ is an $m$-dimensional copula. Denote by $C_S$ the marginal copula function of the subset of variables indexed by $S \subseteq \{1,\dots,m\}$ (e.g.\ $C_{\{1,\dots,m\}} = C$ and $C_{\{i\}}=\text{id}_{[0,1]}$) and by $\bar{C}_S$ the survival function of $C_S$. Similarly, denote by  $\vu_S = (u_i,\ i \in S)$ the vector $\vu \geq \boldsymbol{0} \in \R^m$ indexed by $S$ and let $S_1,S_2\subseteq \{1,\dots,m\}$. The upper tail dependence, exponent and conditional tail dependence functions of $C$ are defined by 
\begin{align*}
	b^*_S(\vu;C) \quad &\overset{\Delta}{=} \quad  \lim_{x \downarrow 0} \quad x^{-1}\bar{C}_S(\boldsymbol{1}-x\vu_S), && \text{(upper tail dependence)}\\
	a^*(\vu;C)\quad   &\overset{\Delta}{=} \quad  \sum_{\substack{S \subseteq \{1,\dots,m\}:\\ S \neq \emptyset}} (-1)^{|S|-1}b^*_S(\vu_S; C), &&\text{(upper exponent)}
\end{align*}

In particular, if $S=\{j\}$ for some $j \in \{1,\dots, m\}$, we define $b^*_{j}(\vu;C) = b^*_{\{j\}}(\vu;C) = u_j$. 
Note that if $C$ has continuous second-order partial derivatives, we have $$a^*(\vu;C) = \lim_{x\downarrow 0} x^{-1}\proba\left(U_i \geq 1-x u_i, \text{ for some $i\in I$} \right),$$
by the inclusion-exclusion principle.

\subsection{Multivariate Generalised Pareto distribution}
\label{section:mgpd}
We suppose that, for any $t \in \N$, the distribution function of $(X^{i}_t,\ i \in I)$ is in the domain of attraction of a multivariate non-degenerate distribution $G_t = G$, written $F \in \mathcal{D}(G)$, if there are vectors $\boldsymbol{a}_n > \boldsymbol{0} \in \R^{d}$, $\boldsymbol{b}_n \in \R^{d}$ and $n \in \N$ such that 
$$F^n(\boldsymbol{a}_n \boldsymbol{x} + \boldsymbol{b}_n) \rightarrow G(\boldsymbol{x}), \quad n \rightarrow \infty,$$
for every continuity point $\boldsymbol{x}\in\R^{d}$ of $G$ where the vector operations and inequalities are meant componentwise. Note that $G$ is necessarily max-stable, i.e.\ there exist vectors $\boldsymbol{a}_n > \boldsymbol{0} \in \R^{d}$, $\boldsymbol{b}_n \in \R^{d}$ and $n \in \N$ such that $G^n(\boldsymbol{a}_n \boldsymbol{x} + \boldsymbol{b}_n) = G(\boldsymbol{x})$ for any $\boldsymbol{x} \in \R^{d}$. By Sklar's theorem \citep{sklar1959fonctions}, there exists a function $C:[0,1]^d \rightarrow [0,1]$ such that
 $$F(x^{1},\dots,x^{d}) = C\left(F^{1}(x^{1}), \dots, F^{d}(x^{d})\right), \quad \text{for all } \boldsymbol{x} \in \R^d, $$
 and $C$ is called the \emph{copula} of $F$. Also, if $F$ is continuous then $C$ is unique.

We suppose that the distribution of the $k$-stack $(\XX_{t}(k), \ t \in \N)$ is in the $\delta$-neighbourhood of a multivariate generalised Pareto distribution (\MGPD) \citep[Definition 1]{falk2019GeneralizedParetoCopulas} which entails 

\begin{enumerate}[label=(\roman*)]
\item marginal distributions of $\XX$ have GPD upper tails;
\item there exists a generalised Pareto copula (GPC) $C_D$ modelling the joint extremes of $\XX_t(k)$;
\item $C$ is in the $\delta$-neighbourhood of $C_D$ for some $\delta > 0$;
\end{enumerate}
which we formalise in the following sections. Alternative \MGPD\ formulations include generator-based distributions \citep{rootzen2018MultivariatePoTModels, kiriliouk2019PoTMultivariateGPD} where extremes events happen when at least one of the components is above a high threshold. 

\subsubsection{Marginal upper tails}
\label{section:marginal-upper-tails}
According to the peaks-over-threshold methodology \citep{davison1990models}, for any $i,j \in I$, we suppose that there are thresholds $\mu^{i,j}_\factual(l)$ and $\mu^{i,j}_\counterfactual(l)$ in $\R$ sufficiently high such that
 \begin{align*}
 	\ermX^{j}_{l}\ &|\ \ermX^{j}_{l} > \mu^{i,j}_\factual(l), \ \mC^{i}  \ \sim \GPD\left(\mu^{i,j}_{\factual}(l),\gamma^{i,j}_\factual(l), \sigma^{i,j}_\factual(l)\right), &&l\in\N\cup\{0\},  &&&\text{(factual)}\\
 	\ermX^{j}_{l}\ &|\ \ermX^{j}_{l} > \mu^{i,j}_\counterfactual(l),\ \mC^{i,c} \sim \GPD\left(\mu^{i,j}_\counterfactual(l),\gamma^{i,j}_\counterfactual(l), \sigma^{i,j}_\counterfactual(l)\right), &&l\in\N,  &&&\text{(counterfactual)}
 \end{align*}
 with shape parameters $\gamma^{i,j}_\factual(l), \gamma^{i,j}_\counterfactual(l) \in \R$ and scale parameters $\sigma^{i,j}_\factual(l), \sigma^{i,j}_\counterfactual(l)>0$. We denote by 
\begin{align*}
	\vmu^{i}_\factual(l) \overset{\Delta}{=} (\mu^{i,1}_\factual(l), \dots, \mu^{i,d}_\factual(l))^{\top} \in \R^d, \quad \text{and} \quad \vmu^{i}_\counterfactual(l) \overset{\Delta}{=} (\mu^{i,1}_\counterfactual(l), \dots, \mu^{i,d}_\counterfactual(l))^{\top} \in \R^d,
\end{align*}
 the vector of factual and counterfactual extreme thresholds. The quantile thresholds and vectors are defined by 
 \begin{equation}
 	\label{eq:quantile-thresholds}
 	\mu^{i,j}_{0,\cdot}(l) \overset{\Delta}{=} \proba(X^j_l \leq \mu^{i,j}_{\cdot}(l)\ | \ \mC^i) \quad \text{and} \quad \vmu^{i}_{0,\cdot}(l) \overset{\Delta}{=} (\mu^{i,1}_{0,\cdot}(l), \dots, \mu^{i,d}_{0,\cdot}(l))^{\top}.
 \end{equation}
 

 \subsubsection{Generalised Pareto Copulas} \cite{deheuvels1984} \& \cite{galambos1987asymptoticExtremeOrder} showed that $F \in \mathcal{D}(G)$ if and only if it is the case marginally, i.e.\ $F^i \in \mathcal{D}(G^i)$, as well as for the copula $C$. The latter is equivalent to $C^n(\vu^{1 / n}) \rightarrow G^*(- \log \vu)$, as $n \rightarrow \infty$, where $ \vu \in (0,1]^m$ and $G^*$ is a max-stable distribution with standard negative exponential margins. Th.\ 2.3.3, \cite{falk2019dnorms} yields that this class of distributions can be formulated as follows
 $$G^*(\vx)  = \exp\left(-\|\vx\|_D\right), \quad \vx \in \R^m,$$
 where $\|\cdot\|_D$ is a $D$-norm ($D$ for dependence) \citep{dehaan2977limitTheoryExtremes, pickands1989negativeExponential} which are characterised by a \emph{generator}, a random vector $(Z^1,\dots,Z^{m})^\top \in \R^m$, such that $Z^i \geq 0$, $\E(Z^i) = 1$ for $i\in \{1,\dots, m\}$. The corresponding D-norm is defined by $\|\boldsymbol{x}\|_D\overset{\Delta}{=}\E\left[\max_{1 \leq i \leq m} (|x^i|Z^i)\right],$ for any $\boldsymbol{x} = (x^1,\dots,x^{m})\in \R^{m}$. In this context, a copula $C_D$ is a GPC if there is a D-norm $\|\cdot\|_D$ on $\R^{m}$ and a vector $\vu_0 \in [0,1)^{m}$ such that
$$C_D(\vu) = 1 - \|\boldsymbol{1} - \vu\|_D, \quad \vu\in [\vu_0, \boldsymbol{1}].$$

\subsubsection{Copulas in a $\delta$-neighbourhood of a GPC}
A copula C is in the $\delta$-neighbourhood of a GPC $C_D$ with D-norm $\|\cdot\|_D$ on $\R^m$ and $\delta > 0$ if the upper tails are closed to one another \citep[Section 6]{falk2019GeneralizedParetoCopulas}, that is if
$$1-C(\vu) =(1-C_D(\vu))\left(1 + O(\|\boldsymbol{1} - \vu\|^\delta)\right)=\|\boldsymbol{1} - \vu\|_D\left(1 + O(\|\boldsymbol{1} - \vu\|^\delta)\right),$$
as $\vu \rightarrow \boldsymbol{1} \in \R^{m}$, uniformly for $\boldsymbol{u} \in [0,1]^{m}$, where $\|\cdot \|$ is an arbitrary norm on $\R^{m}$ (e.g.\ the Euclidean norm).


\subsubsection{The intuition behind the $\delta$-neighbourhood of a GPC}
\label{section:intuition-delta-neighbourhood-gpc}
For a $d$-dimensional copula with standard uniform margins $C$, its upper extreme value (UEV) copula is given by
$$C^{\UEV}(\vu) = \lim_{n\rightarrow \infty} C^n(\vu^{1/n}) = \exp\left\{-a^*(-\log \vu; C)\right\},$$
for any $\vu \in (0,1]^d$ where $a^*(\cdot;C)$ is the upper exponent function of $C$. For instance, both the Archimedean copula and D-vine defined in Prop.\ 4.4 \& 4.5, \cite{joe2010tail} share the same UEV copula which is a GPC \citep[Example 1]{aulbach2019testGPC}. More generally, the upper tail dependence function (and its dual the upper exponent function) characterises the extremal dependance structure of a multivariate distribution \citep[Section 6]{joe2010tail}. If $C$ in the $\delta$-neighbourhood of a GPC, then $C$ converges its UEV copula at a polynomial rate \citep[Th.\ 5.5.5]{falk2011theoremPolynomialGPC}, that is, for some $\delta > 0$, we have
$$\sup_{\vu\in (0,1]^d}\left|C^n\left(\vu^{1/n}\right) - C^{\UEV}(\vu)\right| = O(n^{-\delta}), \quad \text{as $n \rightarrow \infty$.}$$
 The reverse implication is also true under some additional differentiability conditions \cite[Th.\ 5.5.5]{falk2011theoremPolynomialGPC}. Therefore, we assume that a copula is in a GPC $\delta$-neighbourhood if it has a polynomial rate of convergence of maxima \citep[p.\ 602]{aulbach2019testGPC}.
  
 
 \subsubsection{Multivariate regular variation property}
 \label{section:mrv}

A sufficient condition for the limits involved in $a^*$, $b^*$ and $\chi^{(i,j)}(h)$ to exist is that $\XX$ satisfies the multivariate regular variation (MRV) property.
The existence of an upper tail dependence function for a copula $C$ is a necessary and sufficient condition for random vectors with regularly varying univariate marginals to satisfy the MRV property \citep[Th.\ 3.2]{weng2012characterisationMRV}. Many  Archimedean copulas are regularly varying \citep[Table 4.1]{NelsenRogerB2006Aitc} including those used herein (Independent, Clayton, Gumbel, Frank and Joe, see App.\ \ref{section:intro_vine_copulas}). In turn, they have MRV tails \citep[Example 3.5]{weng2012characterisationMRV}. Hence, although the MRV property may not hold for $\XX_t(k) \in \R^{(k+1)d}$ in general, an Archimedean vine copula (App.\ \ref{section:archimedean-regular-vine-copulas}) can be a suitable approximation of the copula of $\XX_t(k)$ that also satisfies the MRV property on $[0,1]^{(k+1)d}$ \citep{weng2012characterisationMRV}.

Note that the MRV property is often assumed without formal checking and this shortcoming has been addressed recently in \cite{einnmahl2020testingMRV} that introduces the first MRV goodness-of-fit test. 

\subsubsection{Copulas and tail dependence functions}
\label{section:tail-dependence-function}
Achieving either only pairwise asymptotic dependence or independence is a limitation of most \MGPD\ formulations \citep{kiriliouk2019PoTMultivariateGPD, ledford1996nearindependence, rootzen2018MultivariatePoTModels, wadsworth2017ModellingAcrossDependenceClasses, huser2019UnknownDependenceClass}. Similarly to the conditional extremes model of \citep{heffernan2004conditional} and recent Markov chain-based models \citep{Winter2017, tendijck2019model, winter2016modelling}, we leverage stationary Archimedean vine copulas that handle both pairwise asymptotic (in)dependence regimes as explained on page 265, \cite{joe2010tail}.

A bivariate copula $C$ is said to verify the (upper) asymptotic linear condition 
\begin{equation}
\label{eq:asymptotic-linear-condition}
	\partial \bar{C}(u,v) / \partial v \approx u \cdot c(v),\quad \text{as $u \rightarrow 0$},
\end{equation}
 for a positive continuous bounded function $c$ \citep[Eq.\ 4.7]{joe2010tail}, which implies the (upper) tail independence of $C$, e.g.\ the independence copula $C(u,v)=uv$. 
 
 Vine copulas have a recursive tree structure where bivariate dependencies are expressed using a stack of conditioned pair-copulas. All pair-copulas presented in Appendix \ref{section:intro_vine_copulas}  have continuous second-order derivatives in $(u,v)$ and only the independence pair copula satisfies (\ref{eq:asymptotic-linear-condition}). Therefore, the upper tail dependence functions are obtained recursively by going up through the vine copula trees \citep[Th.\ 4.1]{joe2010tail}. Also, if the stack of conditioned pair-copulas between two variables has conditional tail dependence functions that are proper distributions and does not feature the independence copula, then those variables are tail dependent \citep[Prop.\ 4.2]{joe2010tail}. 

 Given those notations, we can now link the $\delta$-neighbourhood of a GPC with the upper exponent function of a copula as described in the section below.

\subsection{An alternative to the Markov chain approach} Markov Chains have been a popular tool to implement time clustering of extremes in the last two decades. \cite{smith1997markov} uses a first-order Markov Chain under the assumption of that only asymptotic dependence is possible at lag 1 hence at all lags which is a severe constraint. This limitation was relaxed using $k$-th order Markov Chain \citep{ribatet2009modeling, yun2000distributions} under strict asymptotic dependence. Those limitations are recently been lifted to allow for both asymptotic dependence and independence \citep{Winter2017, tendijck2019model}. Stationary vines \citep{nagler2020stationary} is a multivariate extension of the copula formulation for stationary Markov processes given in Section 2, \cite{Winter2017}. The main difference is that we transform back the vine copula data into the observation scale using unit exponential margins  \citep{wadsworth2013newRepresentationTailProbabilities} to obtain cross-sectionally comparable variables as opposed to Laplace margins in the afore-mentioned article.

\section{Inference}
\label{section:inference}
Suppose we observe  $N$ $d$-dimensional samples $\tX_1,\dots, \tX_{N}$ where $\tX_t = (\etX^1_t, \dots, \etX^d_t)^\top$ and that we are interested in the stack depth of $k \in \N$. Define the factual and counterfactual time sets $I_\factual(i;k) := \{t: \etX^{i}_t > \mu^{i},\ t \leq N - k\}$ and $I_\counterfactual(i;k) := \{t: \etX^{i}_t \leq \mu^{i}, \ t \leq N-k\}$ with the cardinalities $N_\factual = \card(I_\factual)$ and $N_\counterfactual = \card(I_\counterfactual)$.

\subsection{Empirical probability estimates}
 We define the empirical equivalent to the (counter)factual probabilities given in (\ref{eq:linear-factual})  and (\ref{eq:linear-counterfactual}) by
 \begin{equation}
 \begin{aligned}
 \label{eq:empirical-factual-counterfactual}
 \widehat{p}^{(i)}_\factual(v, \vw; k) &\overset{\Delta}{=} N^{-1}_\factual \sum_{t \in I_\factual(i;k)}  \indicator\left\{\vw^\top \tX_{t+1}(k) > v  \right\},\\ 
 \widehat{p}^{(i)}_\counterfactual(v, \vw; k) &\overset{\Delta}{=} N_\counterfactual^{-1}\sum_{t\in I_\counterfactual(i;k)} \indicator\left\{\vw^\top \tX_{t+1}(k) > v \right\},
\end{aligned}
\end{equation}
and we denote the empirical estimate of $\PC$ by $\widehat{\PC}^{(i)}$ for $\PC \in \{\PN, \PS, \PNS\}$, see the simulation study in supplement of \cite{kiriliouk2020ClimateExtremeEventAttribution} about the performance of this estimation approach.
\subsubsection{Objectives}
\label{section:objectives}
In the EEP framework, we quantify the impact of a known marginal extreme event on other variables at subsequent times and, as such, we introduce an optimisation scheme that generalises the idea of \cite{kiriliouk2020ClimateExtremeEventAttribution} to other PCs.

 The weights translate the marginal importance of the components of $\XX_{t+1}(k)$: if the weights are uniform, i.e.\ $\vw = (kd)^{-1}\boldsymbol{1}$,  then no particular variable or time dominates the formation of the impact event. On the other hand, if $\vw$ is such that $w_{d(j-1)+h} = 1$ for some $j\in I$, $h \in \{1,\dots,k\}$ and zero otherwise, then only $X^{j}_{h}$ contributes to the PC value. Weights are either given by an external source such as experts \citep{bevacqua2017multivariate} or derived using the data. We are interested in finding the weight vector that maximises a given $\PC$, which we denote by
$$
	\widehat{\vw}_\PC \overset{\Delta}{=} \argmax \left\{\widehat{\PC}^{(i)}(v,\vw;k): \vw \geq \boldsymbol{0} \in \R^{kd},\ \boldsymbol{1}^\top \vw = 1 \right\}, \quad \PC \in \left\{\PN, \PS, \PNS\right\}.
$$
 We model marginals semiparametrically by leveraging the empirical distribution as well as estimating GPD parameters. After applying the probability integral transform to the data, we fit a unique Archimedean stationary vine copula (ASVC) that captures non-linear dependencies by minimising information criteria such as AIC, BIC and/or mBICV (see Section \ref{section:high-dims}), as indicated below in Section \ref{section:algorithm}.
 \begin{remark}
 	 We expect that dependencies can potentially be different depending on the marginal extreme event $\mC^i_t$ on which one conditions. Therefore, another strategy could be to calibrate different pairs of vine copulas for different marginal extreme events and specifically for factual and counterfactual sub-datasets or for specific time horizons $k$, potentially by avoiding to include all intermediate times $1 \leq l \leq k$.
 \end{remark}

\subsubsection{Regularisation}
\label{section:regularisation}
To extract the most important potential causal links, we also define the weights corresponding to the Lasso or Ridge-type regularisation of the  $\PC$ maximisation  
\begin{equation}
	\label{eq:reg}
		\widehat{\vw}^p_\PC(\lambda) \overset{\Delta}{=} \argmax \left\{\widehat{\PC}^{(i)}(v,\vw;k) - \lambda \|\vw\|_p: \vw \geq \boldsymbol{0} \in \R^{kd},\ \boldsymbol{1}^\top \vw = 1 \right\}, 
\end{equation}
where $p \in \{1,2\}$ and $\lambda \geq 0$ such that $\widehat{\vw}^p_\PC(0) = \widehat{\vw}_\PC$.

 We present the practical inference strategy used to fit both the marginal distributions as well as the ASVC. 

\subsection{Marginal distributions}
\label{section:marginal-distributions}
We consider a simpler model formulation where we do not consider different marginal distributions in the factual and counterfactual worlds. We suppose that
$$
	\vmu = \vmu^{i}_\factual(l) = \vmu^{i}_\counterfactual(l),\quad 
	\gamma^{j} = \gamma^{i,j}_\factual(l) = \gamma^{i,j}_\counterfactual(l), \quad \sigma^{j} = \sigma^{i,j}_\factual(l) = \sigma^{i,j}_\counterfactual(l),
$$
for any $i \in I$ and $l \in \{1,\dots,k\}$ such that thresholds and marginal parameters are independent from the time horizon and the cause on which we condition. Also, the thresholds are independent from the factual or counterfactual marginal distributions such that $\vmu$ coincides to the threshold of $\mC^{i}_t$ defined in (\ref{eq:marginal-extreme-event}). 
 
 We take a semiparametric transformation approach \citep{coles1991modellingExtremeEvents} using an empirical distribution function below the extreme thresholds and perform \GPDTextwithspace maximum likelihood estimation on each marginal independently to obtain estimates for $(\gamma^{j},\sigma^{j})$. We describe the two steps necessary to model the marginal distributions:
\subsubsection{GPD threshold selection} Methodologies to find extreme threshold range from leveraging graphical and diagnostic techniques \citep{davison1990models} to automated selection schemes \citep{bader2018automated, solari2017automaticthreshold}. A threshold that is too low then the GPD approximation for the upper tail may not hold as well and can cause some bias. If it is too high, the sample size is reduced and, in turn, the parameter estimates may have high sample variance. We use the approach described in \cite{bader2018automated} which leverages sequential GPD goodness-of-fit tests on $\etX^{i}_t\ |\ \etX^{i}_t > \mu^{i}$ for a collection of candidate thresholds $\mu^{i}_1<\dots<\mu^{i}_m$ while controlling the false discovery rate on the corresponding sequence of null hypotheses
$$H^{i}_0(l):`` X^{i}\ |\ X^{i} >  \mu^{i}_l \text{ follows a GPD}", \quad l \in \{1,\dots,m\}.$$ 
See \cite{bader2018automated} for a review of alternative methodologies, a comparison of different GoF tests and practical details. Also, we use the implementation from the R package \texttt{eva} \citep{evaPackage2020}.

\subsubsection{GPD parameters} Then, the estimators of the GPD parameters $(\widehat{\gamma}^{i,j}_\factual,\widehat{\sigma}^{i,j}_\factual)$ for any $j \in I$ are obtained by conditioning the data on $\mC^{i}$ and maximising the GPD likelihood of  
$$\left\{ \etX^{j}_{t+l} : \etX^{j}_{t+l} > \mu^{j},\ t \in I_\factual(i;k),\  l \in \{0,\dots,k\}\right\}.$$
This is a special case of composite likelihood maximisation \citep{varin2008composite, varin2011overview}. Even with serially-dependent data, estimators have been shown to convergence asymptotically as the sample size increases \citep{courgeau2021inference}.
Respectively, we estimate $(\widehat{\gamma}^{i,j}_\counterfactual,\widehat{\sigma}^{i,j}_\counterfactual)$ for any $j \in I$ by conditioning on the complement $\mC^{i,c}$ and maximising the GPD likelihood of 
$$\left\{ \etX^{j}_{t+l} : \etX^{j}_{t+l} > \mu^{j},\ t \in I_\counterfactual(i;k), \  l \in \{1,\dots,k\}\right\}.$$
\subsubsection{Semiparametric integral-transform data} Next, the samples are transformed marginally using a semiparametric approach for $t \in \{1,\dots,N\}$:
\begin{align*}
	\etU^j_t
=
\begin{cases}
\widehat{F}^{j}(\etX^j_{t}), \quad &\text{if $\etX^i_{t} \leq \mu^{i}$},\\
\widehat{F}^{j}(\mu^{j}) + \left(1-\widehat{F}^{j}(\mu^{j})\right) \bar{F}_\GPD(\etX^j_{t};\mu^{j},\widehat{\gamma}^{j},\widehat{\sigma}^{j}), \quad &\text{if $\etX^i_{t} > \mu^{i}$},\\
\end{cases}
\end{align*}
where $\widehat{F}^{j}$ is the empirical cdf of $X^{j}$ and $\bar{F}_\GPD(x;\mu,\gamma,\sigma) = (1+\gamma/\sigma(x - \mu))^{-1/\gamma}_+$, the survival function of the GPD cdf.  Another possibility is to use the empirical distribution function throughout; see \cite{shih1995semiparametricCopulaTransfo} for a comparison study.

\subsection{Copula modelling}

Modelling the copula of random vectors to study extremes values is an active research area \citep{bevacqua2017multivariate, falk2019GeneralizedParetoCopulas}. With marginally-uniform data, we fit a unique ASVC on $\tU = (\etU^1_t,\dots,\etU^d_t)^\top$. Another possibility would be to fit two separate vine copulas on factual and counterfactual worlds.

\subsubsection{Vine structure}
\label{section:dissmann-vine-structure}

Selecting the best vine structure is a challenging problem to solve exactly and heuristics are usually used to reduce both the dimension of the vine as well as to describe the tree structures. We describe two approaches, namely the Markov property for vine copulas and a sequential structure selection algorithm from \cite{dissmann2013selecting}.

\paragraph{Markovian vine copulas} 
 In general, a regular vine copula requires fitting $O(N^2d^2)$ pair-copulas (e.g.\ $124,750$ distinct pair-copulas for $N=100$ and $d=5$). The Markov property conveniently reduces the required number of pair-copulas to specify. Formally, a time series $\XX\in\R^d$ is called Markov process of order $p$ if for all $\vx\in \R^d$,
$$\proba\left(\XX_t \leq \vx\ |\ \XX_{t-1},\dots,\XX_1\right) = \proba\left(\XX_t \leq \vx\ |\ \XX_{t-1},\dots,\XX_{t-p}\right).$$
Under this assumption it would require to fit $O(pd2)$ pair-copulas (e.g.\ $60$ pair-copulas for  $p=2$ with $d=5$ and $N=100$).

\paragraph{Reducing the computational complexity} For conciseness, the formal definition of regular vines is relegated to Appendix \ref{section:archimedean-regular-vine-copulas}.
 We perform three tasks when fitting a regular vine: (a) select the vine structure, i.e.\ pick the sets of unconditioned and conditioned pairs of random variables; (b) choose the bivariate copula family; (c) estimate copula parameters. \cite{dissmann2013selecting} proposes a sequential structure selection procedure (Algo.\ 3.1 therein) which maximises the sum of Kendall's $\tau$ \citep{kendall1938new, embrechts2013modelling} of the node pairs given by the edges whilst penalising the total number of edges over the set of all possible spanning trees. Other criteria are also possible: Spearman's $\rho$ \citep{spearman1904rho}, maximum correlation coefficient \citep{gebelein1941statistische} and extremal correlations \citep{engelke2020structure}.
  
Once we know the tree structure of the vine, bivariate copula selection is performed pairwise using information criteria, e.g.\ Akaike (AIC) or Bayesian (BIC), or directly by comparing likelihood values over a family of given pair copulas. However, if the (conditional) pair of random variables cannot reject the null hypothesis of the independence test (say, at the $5\%$ significance level), we use the independence copula \citep{genest2007everything}; as implemented in the R package \texttt{rvinecopulib}. See Section 4.4, \cite{czado2010pair} for additional details.

\subsubsection{Model selection}
\label{section:high-dims}
The augmented flexibility of vine copulas can lead to over-fitting, an issue usually solved by using an information criterion (e.g.\ AIC, BIC). For a copula $C$ with parameters $\boldsymbol{\eta}\in \R^m$, we denote by $q_{\max}$ the maximum number non-zero parameters $C$ can have, e.g. for a vine copula with 1-parameter pair-copula $q_{\max} = m(m-1)/2$. In equal likeliness of all vine copulas, if we suppose that the true model has $q$ non-zero dependence parameters (i.e.\ non-independent copulas), then we observe that $q/q_{\max} \rightarrow 1/2$ as $m\rightarrow \infty$ when minimising the BIC. 

To handle this issue, we use a sparse information criterion with a heavier penalty when $q$ grows large and is denoted by \mBICV \citep[Section 3.2]{nagler2018model}. This criterion consists in defining independent Bernoulli random variables for each edge (i.e.\ pair copula) in a vine tree with a prior probability of not being independent, penalising heavily trees with more non-independent copulas and higher in the hierarchy. 
In high-dimensional problems where $q_{\max}=m(m-1)/2$ rivals with $N$, this penalty allows us to find models where $q \ll q_{max}$ or asymptotically, $q/q_{max}\rightarrow 0$ as $m \rightarrow \infty$ \citep[Section 4]{nagler2018model}. This allows to prune the vine of its less significant links.

\subsection{Intervention sampling using the vine copulas}
\label{section:intervention-sampling}
As explained in Section \ref{section:causal-models}, causal models comprise a graph structure that describes the causal (and directed) relationships between the exogenous and endogenous variables. On the other hand, vine copulas are a collection of undirected trees that we use as a generative model. By conditioning on a subset of variables (e.g.\ a particular marginal extreme event), we transform implicitly each undirected tree into its directed ones whose edges point outwards from known variables to the unknown variables. We propagate the known values on which we condition through the trees using an iterative transformation called the inverse Rosenblatt function \citep{rosenblatt1952transform}. Generated samples satisfy the dependence structure described by the vine and we mimic the intervention mechanism described in App.\ \ref{section:time-series-causality}. See \cite{cooke2015sampling} and App.\ B in \cite{bevacqua2017multivariate} for more details on the conditioning of vine copulas.

\subsection{Algorithm}
\label{section:algorithm}
To undertake the model inference as done in Section \ref{section:london-air-pollution-dataset}, we perform the following steps in order:
\begin{enumerate}[label=(\roman*)]
	\item GPD threshold selection using a sequential testing procedure;
	\item Maximum likelihood estimation of GPD tails;
	\item Probability integral transform on each marginal;
	\item Fitting a Markovian stationary Archimedean vine copula using information criteria;
	\item Generate factual and counterfactual data to estimate the tail probabilities;
	\item Compute, and potentially optimise for, the probabilities of causation.
\end{enumerate}

\subsection{Estimating the tail probabilities}
\label{section:tail-probabilities-approx}
In this section, we review options available for applications to estimate the (counter)factual tail probabilities given in (\ref{eq:linear-factual}) and (\ref{eq:linear-counterfactual}) that are essential to computing the probabilities of causation as the impact threshold $v$ gets larger and the sample size gets thinner.
\subsubsection{Approximations available}
The impact random variable $\vw^\top \XX_{t+1}(k)$ have been interpreted in two different ways. First, as a univariate random variable where $\vw$ is kept fixed and where the GPD approximation holds "pointwise" \citep{bevacqua2017multivariate}, a technique that suitable when the weights are known in advance, e.g.\ suggested by experts. Second, as a weighted sum of components of a random vector, where $\vw$ is seen as parameters of the model that modulates the GPD scale parameter \citep{kiriliouk2020ClimateExtremeEventAttribution} while all marginals are assumed to share the same shape parameter $\gamma$. This modulation allows to solve alternative optimisation problems (see Section \ref{section:objectives}). The fixed shape parameter makes this approach better suited for applications where one expects the marginal distributions to be relatively close to one another (e.g.\ precipitation data over multiple weather stations in \cite{kiriliouk2020ClimateExtremeEventAttribution}). They take the average shape parameter as the unique shape parameter which works well in that case but may not be suitable in general. 

\subsubsection{Marginal transformation}In general, the components of $\XX$ can represent very different quantities (e.g.\ air pollutant concentrations and number of cars on a street). To make them comparable in scale beyond a potential unit-variance normalisation, we also consider a modified linear impact function given by
$$h(\vx;\vw) = \sum_{i=1}^d\sum_{j=1}^k w_{d(i-1) + j} H\left(F^i(x^i_j)\right),$$
where $H:\R \rightarrow \R$ is any continuous function and $F^i$ is the cdf of the $i$-th component $X^i$. For instance, the identity $H=\text{id}$ implies that impact event related to the weighted sum of uniformly-distributed values. On the other hand, for $H(v)=F^{-1}_\GPD(v ;\gamma, \sigma)$, then we consider the weighted sum of potentially dependent variables such that $$H(F^i(X^i_j)) \sim \GPDText(\gamma,\sigma).$$
For $\vw^\top \vone = 1$ and $v > \vw^\top \vmu_0$, we have similar tail approximations given by
\begin{align*}
	p^{(i)}_\factual(v, \vw; k) 
	\quad &\approx \quad \widehat{p}^{(i)}_{\factual,H}(\vw^\top H(\vmu_0), \vw; k)\times\widebar{F}_\GPD(v;\vw^\top H(\vmu_0),\gamma,\sigma),\\
	p^{(i)}_{\counterfactual}(v, \vw; k) 
	\quad &\approx \quad \widehat{p}^{(i)}_{\counterfactual,H}(\vw^\top H(\vmu_0), \vw; k)\times\widebar{F}_\GPD(v;\vw^\top H(\vmu_0),\gamma,\sigma),
\end{align*}
where $\widehat{p}^{(i)}_{\factual,H}$ and $\widehat{p}^{(i)}_{\counterfactual,H}$ are computed with respect to the transformed data, that is
\begin{equation*}
 \widehat{p}^{(i)}_{\cdot,H}(v, \vw; k) \overset{\Delta}{=} N^{-1}_\factual \sum_{t \in I_\cdot (i;k)}  \indicator\left\{\vw^\top \cdot H\circ F\left(\tX_{t+1}(k)\right) > v  \right\},
\end{equation*}
where the transformation is componentwise, i.e.\ the $(d-1)i+j$-th component of $H\circ F\left(\tX_{t+1}(k)\right)$ is given by $H(F^i(\etX^i_{t+j}(k))$. We introduce the marginal transformation trick as a potential solution to this issue under the additional constraint that the weighted sums are with respect to the $H$-transformed data and the impact threshold ought to be adapted accordingly.

\section{London air pollution dataset}
\label{section:london-air-pollution-dataset}

The data was provided by King's College London Air Quality Network which provides air pollution data on different timescales and pollutants for many locations in Greater London. All the chemicals reactions mentioned below are described in detail in the related air pollution literature from the World Health Organization \citep{krzyzanowski2008update}. An R implementation of the causal framework can be found on GitHub.\footnote{https://github.com/valcourgeau/xvine} The script used to generate results and plots can be retrieved in another repository.\footnote{https://github.com/valcourgeau/extreme-applications}

Our dataset consists of hourly measurements of the six main air pollutants: Ozone ($O_3$), Nitrate Oxide ($\texttt{NO}$), Nitrate Dioxide ($\texttt{NO2}$), Carbon Oxide ($\texttt{CO}$), Particulate Matter under 10 microns ($\texttt{PM10}$) and Sulphur Dioxide ($\texttt{SO2}$), as well as the total number of vehicles per hour ($\texttt{v/hr}$) from $1^{\mathrm{st}}$ January 2000 to $13^{\mathrm{th}}$ April 2002 on Marylebone road, London, UK. Those ground-level measurements correspond to 20,000 entries and are communicated in $\mu g / m^3$.

We are interested in studying the causal impact of traffic spikes (extremes in \texttt{v/hr}) on air pollutants levels over fix hour window. An Archimedean stationary vine copulas is used the model the dependencies between marginals and compare the probabilities of causation with uniform weights $w_{l} = 1/kd$ for $l \in \{1,\dots,kd\}$. Finally, we maximise the PCs with respect to the weights.

\subsection{Data preparation and exploration}
\label{section:data_prepa_air_pollution}

\begin{figure}[htp]
  \begin{center}
\includegraphics[width=1.0\textwidth]{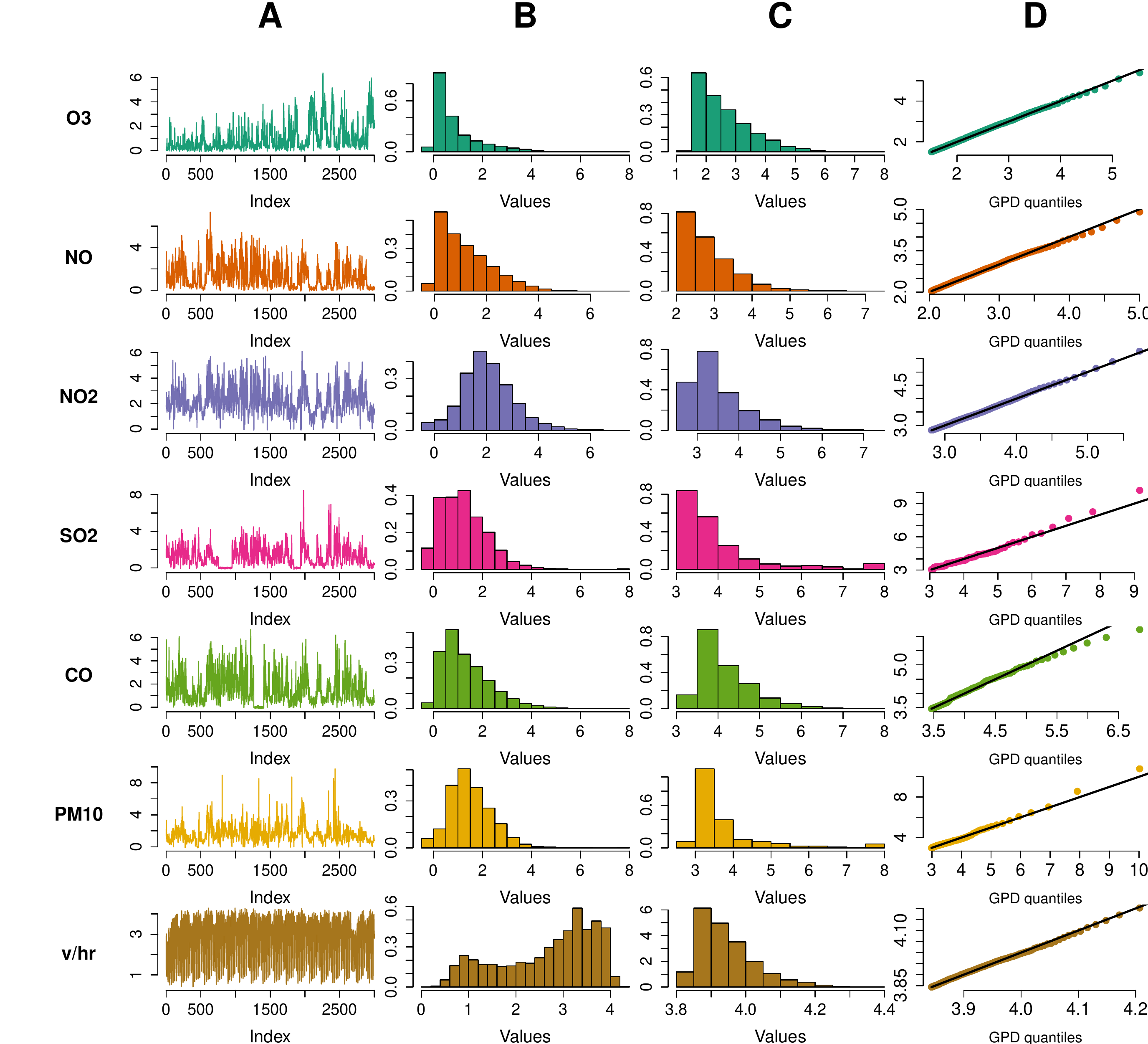}
\caption{(A) Time series (B) Histograms (C) Histogram above extreme threshold (D) GPD Q-Q plot.}
\label{figure:explo}
\end{center}
\end{figure}

To avoid rounding issues from the air pollutants sensors, we jitter the data with Gaussian noise with mean zero and standard deviation equal to 5\% of the original time series standard deviation. All marginals are then normalised to the unit variance and they reject the null hypothesis of the Augmented Dicker-Fuller test with a significance level below $1\%$ suggesting that they are stationary. Marginal histograms, as well as histograms above the extreme threshold and GPD Q-Q plots, can be found in Figure \ref{figure:explo}. We observe that all air pollutant marginals are unimodal distributions (with the mode between 0 and 2) whilst $\texttt{v/hr}$ is slightly bimodal with a skew towards a high traffic mode around $\approx 3.5$. The extreme threshold selection (Section \ref{section:marginal-distributions}) was deployed with a 5\% significance level. We note that thresholds may not be necessary as high as expected since the $80\%$ quantile provide a good GPD fit as shown in \textbf{D}, Fig.\ \ref{figure:explo}. It failed for \texttt{SO2} and \texttt{CO} for which the extreme threshold was set to be $96\%$ quantile.

\begin{figure}[htp]
  \begin{center}
\includegraphics[width=1.0\textwidth]{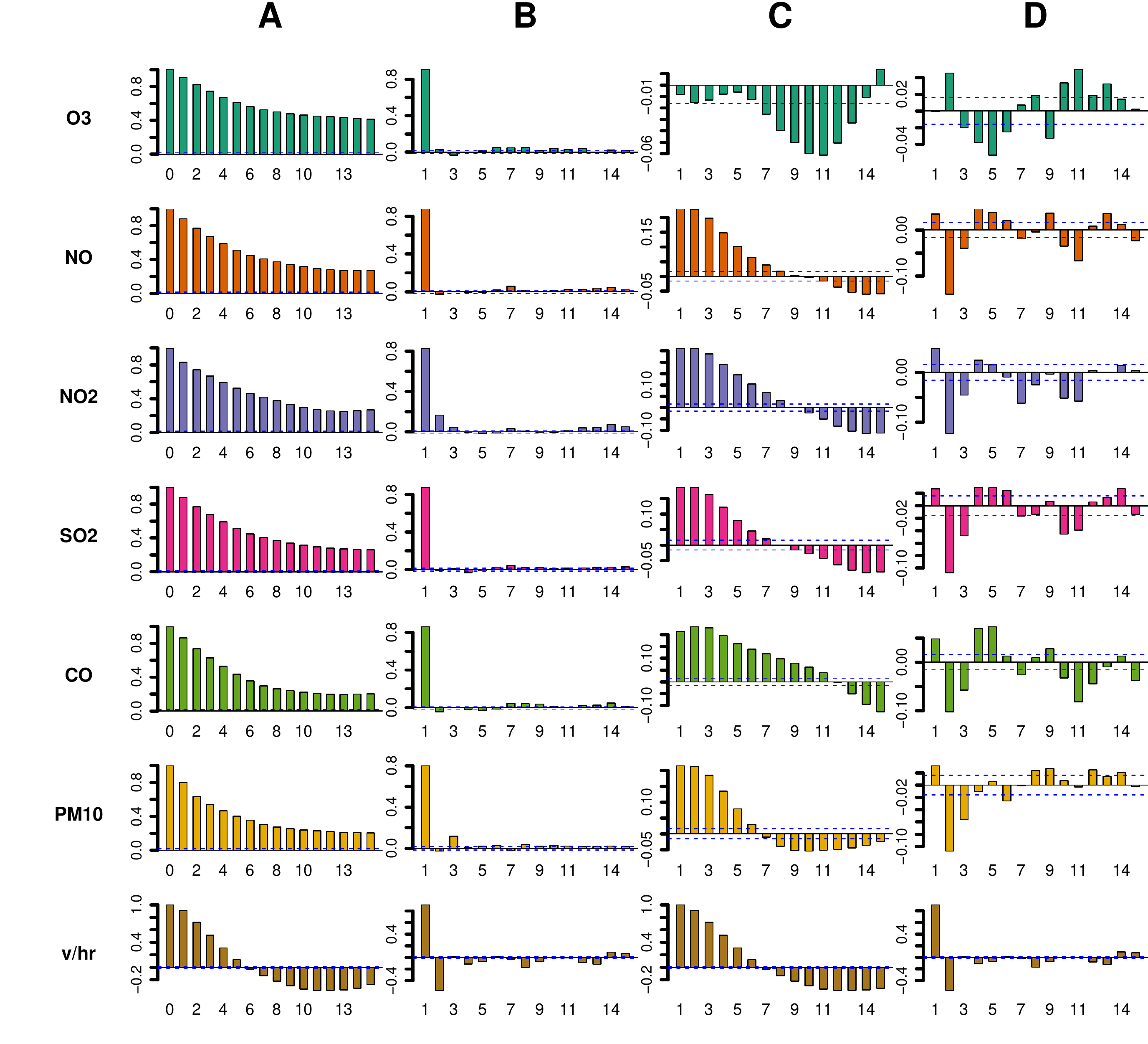}
\caption{Correlation functions as a function of the lag difference: (A) Autocorrelation (B) Partial Autocorrelation (C) Cross-correlation with \texttt{v/hr} (D) Partial cross-correlation with \texttt{v/hr}.}
\label{figure:acf}
\end{center}
\end{figure}

In Figure \ref{figure:acf}, autocorrelation (acf) and cross-correlation (ccf) functions with \texttt{v/hr} (along with corresponding partial quantities pacf and pccf) show that there is autoregressive serial dependence marginally. Pacf values are mostly insignificant after lag 1 except for $\texttt{v/hr}$ where there is a negative pacf at lag 2. For all marginals except $\texttt{O3}$, ccf (resp.\ pccf) values are strongly positive for small lags (resp.\ negative at lag 2) and decrease as the lag increases until they become negative after lags 8--10. For \texttt{O3}, there is a large negative ccf spike around lag 10. Those observations coincide with what is suggested by information criteria (AIC/BIC) of the Markovian vine copulas (Table \ref{table:vine-ic}), see Section \ref{label:vine-fitting-selection-example}.

\begin{table}[htp]
\begin{center}
\begin{tabular}{rccccc}
\specialrule{.1em}{.1em}{.1em}
\multicolumn{1}{l}{} & $\boldsymbol{\gamma}$ & $\boldsymbol{\sigma}$ & $\boldsymbol{\mu}$ & $\boldsymbol{\mu}_0$ & \textbf{ADF stat.} \\
\hline
\texttt{O3} & -0.191 (.010) & 1.316 (.024) & 1.495 & 80\% & -19.193 \\
\texttt{NO} & -0.152 (.009) & 0.899 (.016) & 2.023 & 80\% & -23.746 \\
\texttt{NO2} & -0.073 (.016) & 0.748 (.017) & 2.806 & 80\% & -24.321 \\
\texttt{SO2} & 0.228 (.040) & 0.751 (.040) & 3.041 & 96\% & -23.781 \\
\texttt{CO} & 0.032 (.027) & 0.684 (.030) & 3.454 & 96\% & -26.683 \\
\texttt{PM10} & 0.431 (.039) & 0.484 (.023) & 2.981 & 94\% & -25.398 \\
\texttt{v/hr} & -0.202 (.017) & 0.122 (.004) & 3.842 & 92\% & -56.342 \\
\specialrule{.1em}{.1em}{.1em}
\end{tabular}                    
\label{table:f-cf-parameters}
\caption{Unconditional GPD parameters estimates; extreme levels $\mu$ and corresponding quantile $\mu_0$; Augmented Dickey-Fuller (ADF) test statistic (critical value at the 1\% significance\ level: $-3.43$). Asymptotic standard deviations are in parenthesis when available.} 
\end{center}
\end{table}


\subsection{Vine selection}
\label{label:vine-fitting-selection-example}

By AIC minimisation, we fit an ASVC with a Markov order of $10$ following the works of \cite{nagler2020stationary} with Independence, Clayton, Gumbel, Frank \& Joe pair-copulas, see Appendix \ref{section:archimedean-regular-vine-copulas}, which also coincides with the best order with respect to BIC. We conjecture that this reflects the capacity of the vine to capture the weaker cross-sectional dependencies observed at lags $8-10$. 

\begin{table}[htp] 
\begin{center}
\begin{tabular}{rcccccccccc}
\specialrule{.1em}{.1em}{.1em}          
Order & 1 & 2 & 3 & 4 & 5 & \dots & 9 & 10 & 11 & 12 \\ \hline
AIC & -22641 & -23208 & -23340 & -23361 & -23580 & \dots & -23713 & \textbf{-23839} & -23688 & -23688 \\
BIC & -21994 & -22441 & -22528 & -22534 & -22723 & \dots & -22796 & \textbf{-22877} & -22786 & -22786 \\
mBICV & -22151 & \textbf{-22394} & -22036 & -21295 & -20401 & \dots & -11147 & -7342 & -2346 & 3263 \\
\specialrule{.1em}{.1em}{.1em}          
\end{tabular}
\label{table:vine-ic}

\caption{Information criteria as a function of the ASVC Markov order.} 
\end{center}
\end{table}

On the other hand, the modified criterion $\mBICV$ with $\psi_0 = 0.9$ suggest an order of $2$, presumably only capturing the strong autoregressive serial dependencies suggested by the acf/pacf values (\textbf{A} \& \textbf{B},\ Fig.\ \ref{figure:acf}). 

\subsubsection{Synthetic data}
As mentioned in Sections \ref{section:time-series-causality} \& \ref{section:intervention-sampling}, we leverage the conditional sampling feature of the ASVC to approximate the intervention sampling mechanism. In the 20,000 samples of the dataset, a \texttt{v/hr} marginal extreme event occurs in $8\% \approx 1,500$ of them. Therefore, we generate $1,500$ samples of $\XX_{t+1}(k)$ for $k=6$ given $\XX_t = \vx_t$ such that $x^{\texttt{v/hr}}_t > \mu^{\texttt{v/hr}}$ where $\vx_t$ is taken from the dataset. Similarly, we simulate $1,500$ samples of $\XX_{t+1}(k)$ where $\XX_t = \vx_t$ such that $x^{\texttt{v/hr}}_t \leq \mu^{\texttt{v/hr}}$ (and $\vx_t$ is also from the dataset). Note that we fix the other marginals $j \neq {\texttt{v/hr}}$ to prevent instantaneous causality as given in (\ref{eq:intervention-indicator-assumptions}).

For comparison purposes, we downsample the dataset into two sets of $1,500$ realisations of $\XX_t(k)$ with/without the \texttt{v/hr} marginal extreme event such that $\XX_t = \vx_t$ is the same as the one used in the conditional sampling approach mentioned above. We use those two sets to compute empirical (counter)factual probabilities and corresponding PCs at a similar level of uncertainty as with the vine-generated data.

Finally, we apply a unit Exponential marginal transformation to the data ($\gamma = 0$ and $\sigma = 1$) as mentioned in Section \ref{section:tail-probabilities-approx}. Also, the impact threshold $v\in\R$ is set to the $80\%$-quantile of the unit Exponential, see Section \ref{section:limitations} for a remark on the matter.

\begin{figure}[htp]
  \begin{center}
\includegraphics[width=1\textwidth]{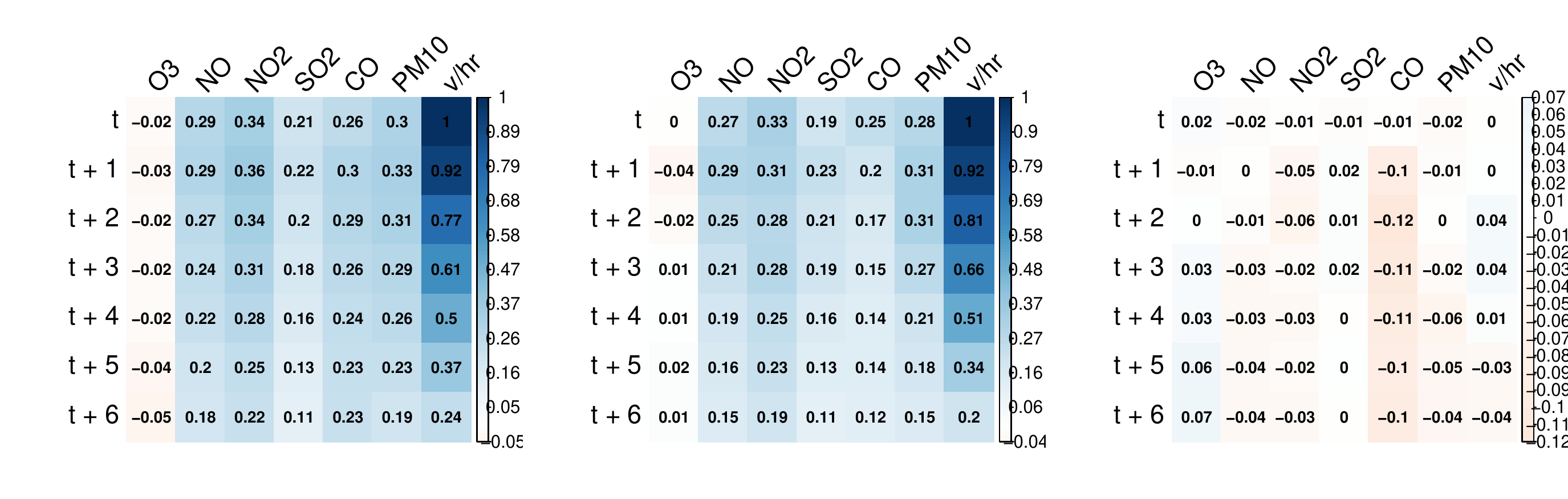}
\caption{Componentwise correlation between $X^\texttt{v/hr}$ and $\XX_{t+1}(k)$ of (left) real data; (middle) synthetic data; (right) their componentwise difference.}
\label{figure:correlation}
\end{center}
\end{figure}
As pictured in Figure \ref{figure:correlation}, the correlation between $X^\texttt{v/hr}$ and other marginals are all positive for $t+1,\dots,t+6$ except for \texttt{O3} which are all not significantly non-zero. Those features are well replicated in the dataset as shown in the middle and right-hand side plots with correlation differences below $6\%$ all components across the following 6-hour window, except for \texttt{CO}. For that particular pollutant, correlations are underestimated in the synthetic data with error ranging from $-10\%$ to $-12\%$.

\subsubsection{Uniform weights}
\begin{figure}[htp]
  \begin{center}
\includegraphics[width=1.\textwidth]{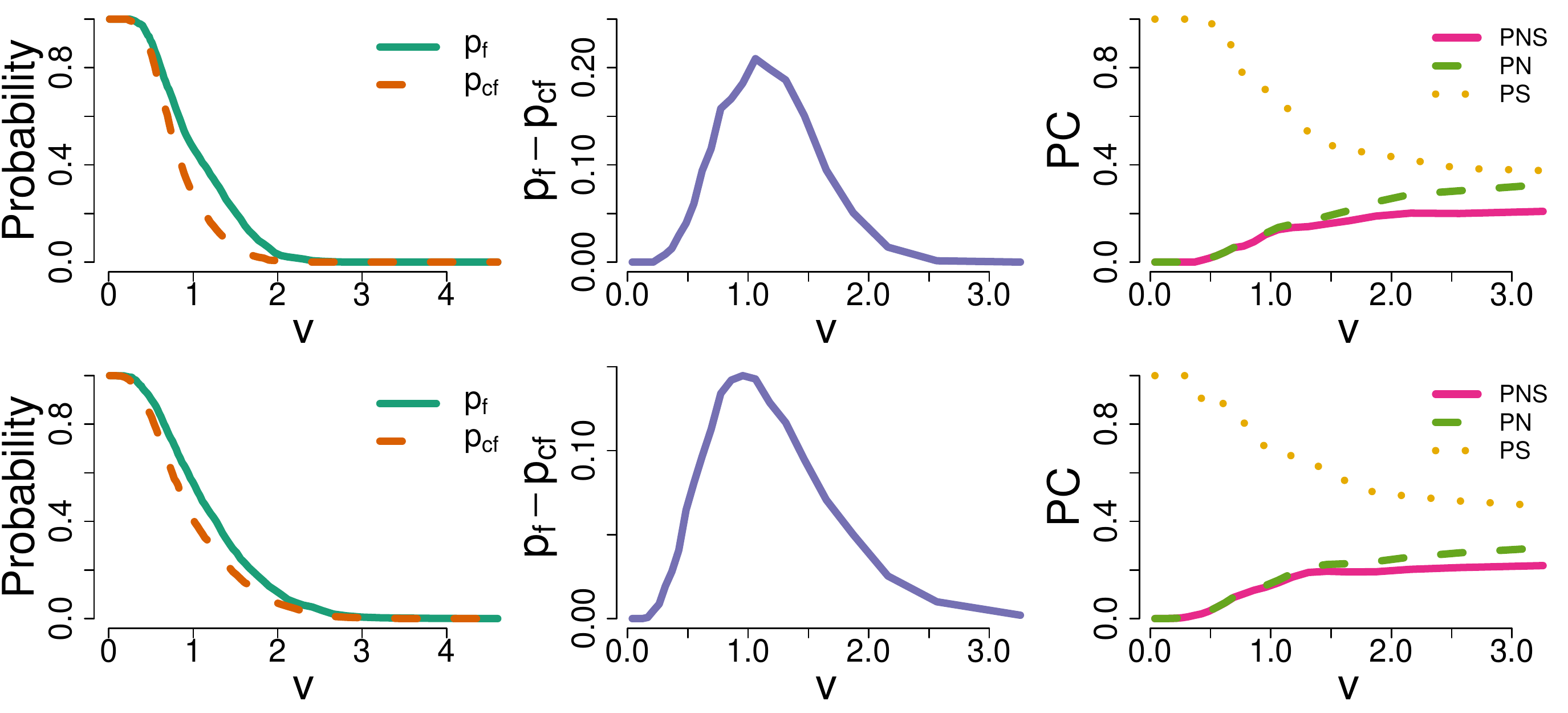}

\caption{Probabilities with uniform weights for (top) real data and (bottom) synthetic data: (left) $p_\factual$ and $p_\counterfactual$ against the impact threshold $v$; (middle) their difference; (right) the corresponding probabilities of causation (PNS in pink bold, PN is green dashed and PS in yellow dotted lines).}
\label{figure:p-f-cf-unif-weights}
\end{center}
\end{figure}

In this section, we focus on the case of a uniform weight vector, that is, where $w_l = 1/kd$ for $l \in \{1,\dots,kd\}$ and no component controls the likelihood of the impact event except through their factual and counterfactual distributions. We show qualitatively in Figure \ref{figure:p-f-cf-unif-weights} that the synthetic data replicates the main characteristics of (counter)factual probabilities and the associated PCs as the impact threshold $v$ grows. 

The fact that $p_\factual > p_\counterfactual$ is a necessary condition for the \emph{monotonicity} assumption as shown on the two left-hand side and middle plots in Fig.\ \ref{figure:p-f-cf-unif-weights}. The empirical (counter)factual probabilities showcase a wider difference $p_\factual - p_\counterfactual$ of $20\%$ for $v=1.2$ as opposed to the maximum of $15\%$ at also $v=1.2$ for the synthetic probabilities. Although the factual probabilities share similar behaviour across the different values of $v$, the synthetic counterfactual probabilities stay higher as opposed to their empirical counterparts. Both probabilities are approximately equal as $v \downarrow 0$ or as $v\uparrow \infty$. The probabilities of causation are very similar too: $\PS$ decreases from $100\%$ to $\approx 60\%$ as $v\rightarrow \infty$ whilst $\PN$ and $\PNS$ increase hand-in-hand from $0\%$ to approximately $\approx 40\%$. Although \cite{kiriliouk2020ClimateExtremeEventAttribution} reported that PNS is not monotonous as a function of $v$ for simulated data but rather bell-shaped, it is increasing in our case. On the other hand, PN values are lower than those stated in the aforementioned article. We attribute those differences to the underlying structure and dependencies of the data.

\subsection{Maximising the probabilities of causation}
In this section, we focus on maximising the PCs with respect to the weight vector $\vw$ (Section \ref{section:objectives}).
\subsubsection{Implementation details}
The optimisation is done in two stages: starting from the uniform weights, we build an initial guess in $[0,1]^{kd}$ using the differential evolution algorithm, a global optimisation routine implemented in the R package \texttt{DEoptim} \citep{mullen2011deoptim} with a number of candidates equal to ten times the dimension of $\vw$. Next we optimise by using the L-BFGS-B scheme, potentially by adding a regularisation term as in (\ref{eq:reg}) and explored in Section \ref{section:regularisation-air-pollution}. We present standardised weight matrices that are computed as follows
$$\tilde{w}_{l} = \frac{w_{l}}{\max_{1 \leq m \leq kd} w_{m}} \in [0,1].$$
Also, we set the threshold $v$ to be the $80$\%th quantile of the unit Exponential distribution, i.e.\ $v \approx 1.60$.

The PNS and PS weight matrices look very similar, with high weights for \texttt{v/hr} for $t+1$ through $t+4$.
\begin{figure}[htp]
  \begin{center}
  \hspace*{-0.5cm}
\includegraphics[width=1.05\textwidth]{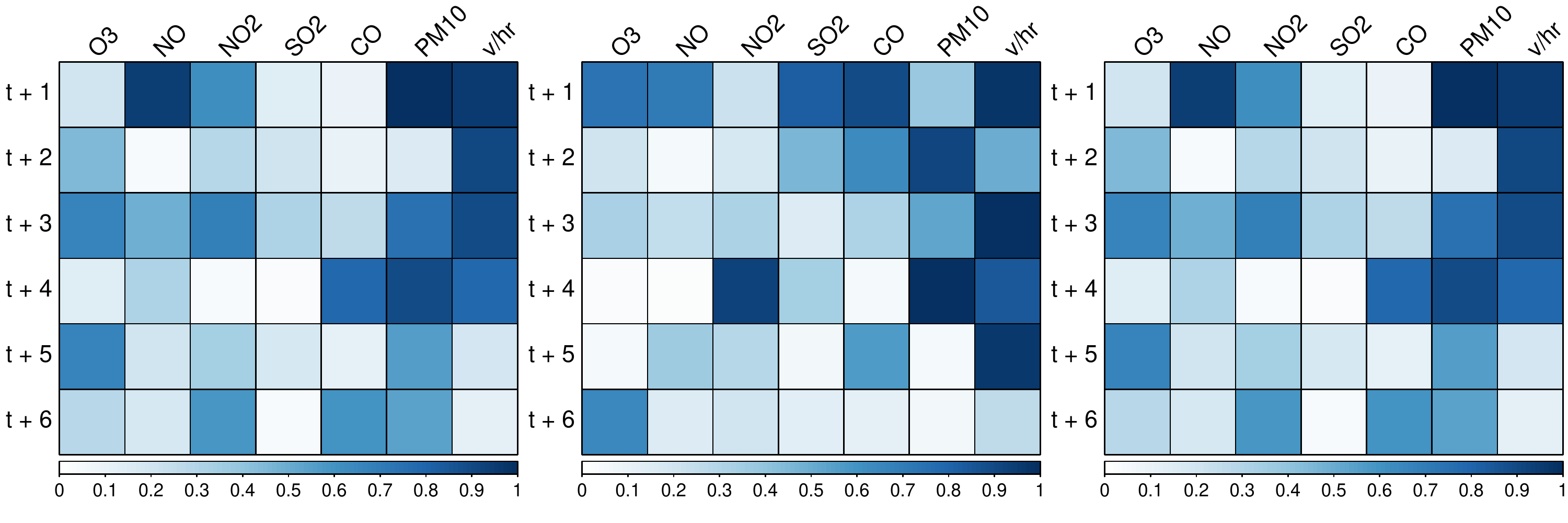}

\caption{Weight matrices after maximisation: (left) PNS; (middle) PN; (right) PS.}
\label{figure:matrices}
\end{center}
\end{figure}
At $t+1$, both the weights of  \texttt{NO2} and \texttt{PM10} are also high which is sensible since they are related to exhaust fumes, road dust and pollen material produced or moved by cars \citep[p.\ 180]{coria2015airpollution}. The \texttt{NO} weight is also very high as it may be produced by the oxidation of \texttt{NO2} \citep[p.\ 444]{crutzen1979role}. We also observe mid-ranged weights at $t+3$ and $t+5$ for \texttt{O3} which may come as a by-product of nitrate oxides (\texttt{NO} \& \texttt{NO2}, see p.\ 445, \cite{crutzen1979role}) or from volatile organic compounds, such as \texttt{CO}, coming from gasoline combustion, that react with ultraviolet rays  \citep[p.\ 445]{crutzen1979role}. We refer the interested reader to the comprehensive review in \cite{crutzen1979role} for a detailed summary of the competing chemical reactions.

Regarding the weights related to PN, we observe that all marginals except \texttt{NO2} are above $40\%$ of the maximum weight at $t+1$. \texttt{CO} has much higher weights than in the PNS and PS cases, with $70\%$ of the maximum at time $t+1$ and between $20\%$ and $40\%$ at times $t+2, t+3$ compared to close to $0\%$ in either the PNS or the PS case.

\subsubsection{Influence of the cause marginal on the impact event}
As pictured in Fig.\ \ref{figure:acf}, \texttt{v/hr} exhibits an autoregressive serial dependence which could be picked up the causal probabilities as a causal link. To verify this statement, we perform an ablation study where we remove the cause marginal (\texttt{v/hr}) from the impact event and observe the potential changes in the PCs and distribution of the weights.
\begin{figure}[htp]
  \begin{center}
\includegraphics[width=1.\textwidth]{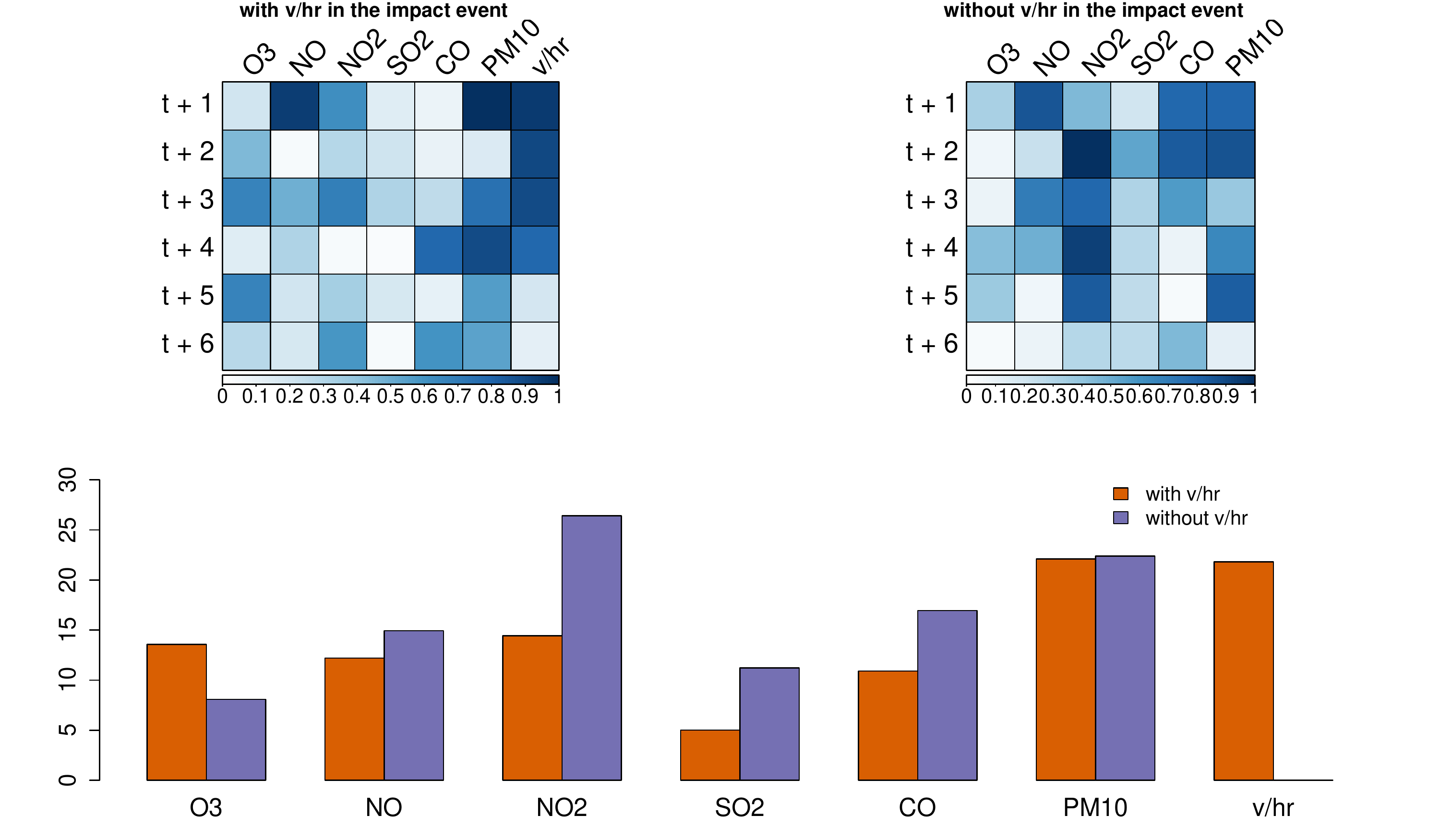}

\caption{PNS maximisation with or without the cause marginal \texttt{v/hr} in the impact event. (top left) standardised weights that include \texttt{v/hr}; (top right) standardised weights that \emph{do not} include \texttt{v/hr}; (bottom) contribution (in \%) of each marginal in the weight vector either \emph{with} \texttt{v/hr} or \emph{without} \texttt{v/hr} in the impact event.}
\label{figure:aws}
\end{center}
\end{figure}

Removing \texttt{v/hr} changes the output substantially. In Figure \ref{figure:aws}, we plot the PNS weight matrices after maximisation with and without \texttt{v/hr} in the linear impact event. The bottom bar plots show the proportion of the weights attributed to each marginal separately. We can infer that most of the weight mass is transferred onto \texttt{NO2} that increases from $15\%$ to more than $25\%$. The weight concentration on \texttt{PM10} remains high around $20\%$, \texttt{SO2} almost doubles to 12\% whilst \texttt{03} is almost halved from $13\%$ down to $7\%$. See Section \ref{section:regularisation-air-pollution} for additional details on the impact of adding the cause marginal into the impact event.

\subsection{Regularisation}
\label{section:regularisation-air-pollution}
Regularisation is a usual tool to increase the weight concentration on the most significant contributing variables. A Ridge-like regularisation ($p=2$) is equivalent to assuming a Gaussian prior on the weights. The case $p=1$ implies a Laplace prior distribution on the weights, which features a higher probability around zero than the Gaussian case. This means that more insignificant weights will most probably be set to zero. 

In this section, we showcase the behaviour of the impact event weights as we increase the regularisation parameter $\lambda$ starting from the unregularised case ($\lambda = 0$). We present the maximal PCs (PNS, PN, PS) obtained for different values of $\lambda$ along with the entropy of their weights, a statistic that we detail in the next section

\subsubsection{Shannon's entropy}
We suppose that the sum of the weights is equal to \emph{one}, i.e.\ $\boldsymbol{1}^\top \vw = 1$. To characterise the distribution of the weights, we compute the standardised Shannon's entropy \citep{shannon1948mathematical} of the impact weights; that is, relative to the uniform weights which maximise this said entropy to obtain a value between $0$ and $1$. The statistic is given by 
$$-\vw^\top \log \vw / \log(kd) \in [0,1].$$ 
The relative entropy is equal to \emph{one} when the weights are uniformly distributed ($w_l = 1/(kd)$ for all $l \in \{1,\dots,kd\}$); and equal to \emph{zero} when the weights are concentrated on a particular component: i.e.\ if for some $j \in \{1,\dots,kd\}$, we have
$$w_l = \begin{cases}
	1,& \text{if $l = j$},\\
	0, &\text{if $l \neq j$}.
\end{cases}$$
Between the two corner cases, the relative entropy gives some information about the concentration of the weights, from the uniformly distributed case to a deterministic distribution. We use this value to quantify the role of the regularisation term on the weight vector distribution after maximisation.

\subsubsection{Results}

We proceed to maximise the PNS, PN and PS with respect to $\vw$ when \texttt{v/hr} is included, or not, in the impact event for an increasing regularisation parameter $\lambda \in \{0, 0.001, 0.01, 0.1, 0.5, 1, 5, 10, 100, 200\}$ and $p\in\{1,2\}$. Again, the impact extreme threshold is set as the $80\%$ Exponential quantile, i.e.\ $v \approx 2.52$.

\begin{figure}[t]
  \begin{center}
\includegraphics[width=1.\textwidth]{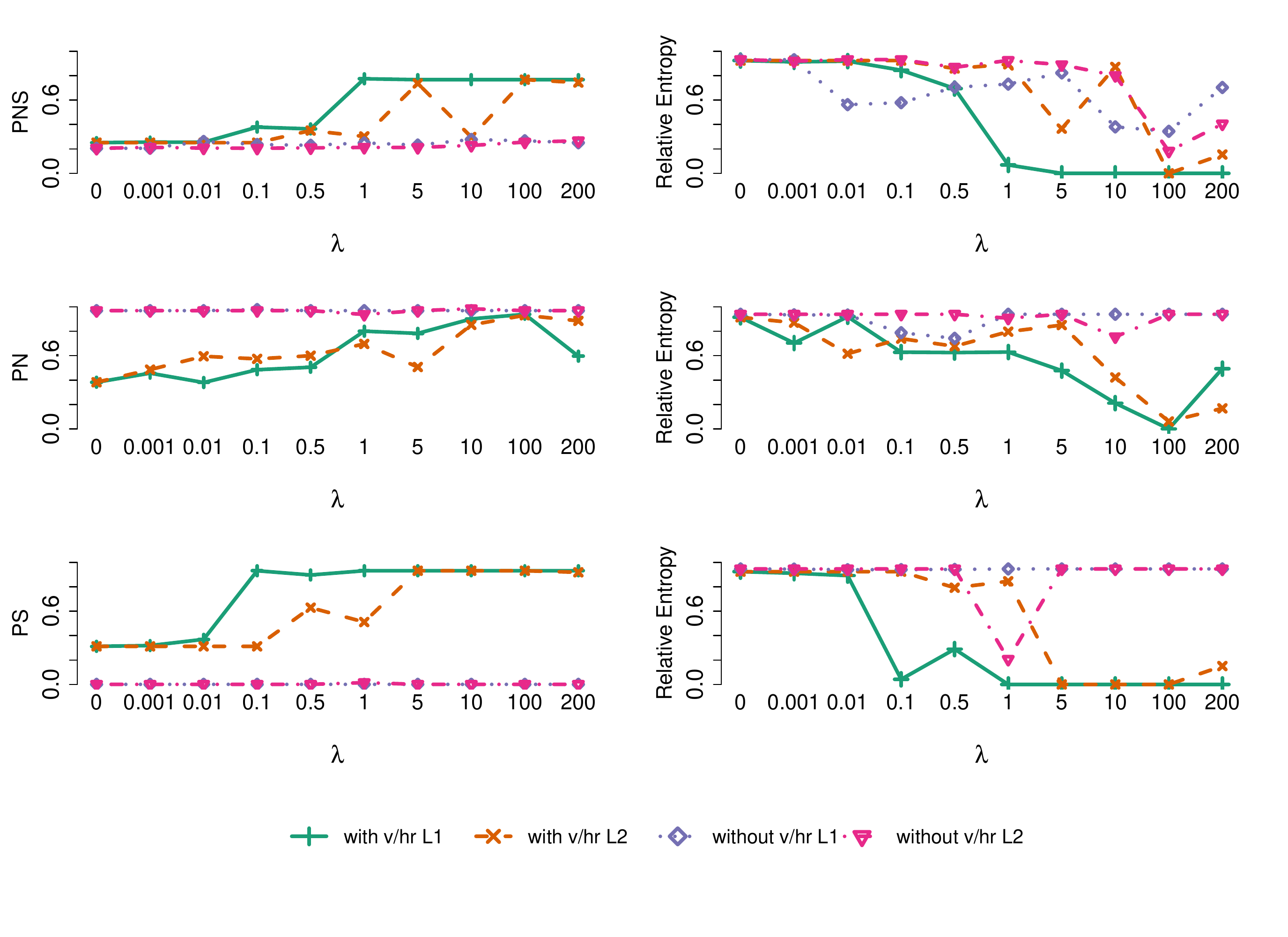}

\caption{PC maximisation as a function of the regularisation parameter $\lambda$: (left) PNS, PN and PS values obtained, respectively. (right) entropy relative to the uniform case (lower means the weights are more concentrated) for each case;}
\label{figure:entropy}
\end{center}
\end{figure}

\paragraph{Probabilities of causation} Figure \ref{figure:entropy} features all three cases with each row containing both the maximised PC value and the relative entropy as a function of $\lambda$. In the first row, the relative entropy of both cases with and without \texttt{v/hr} starts close to $1$ and decreases as $\lambda$ gets larger, down to $0$ for $\lambda \geq 5$ when \texttt{v/hr} is included under an $L^1$ penalty (i.e.\ $p=1$). We observe that the PNS stays relatively constant at $\approx 20\%$ across the four cases for $0 \leq \lambda \leq 0.5$. Then, when \texttt{v/hr} is included under an $L^1$ penalty, as the relative entropy gets lower, the PNS surges past $95\%$ for any $\lambda \geq 1$. Similar behaviour is also obtained for the $L^2$ penalty, albeit decreasing sharply for $\lambda = 10$. We explain this decrease by the fact that the regularisation may change the optimisation landscape in a non-linear manner. On the other hand, when \texttt{v/hr} is not included in the impact event, the PNS stays constant while the relative entropy decreases at about $50\%$ for all three other cases, showing that the regularisation has an effect on the weight vector although not increasing the PNS when \texttt{v/hr} is not included. This indicates that, as expected, \texttt{v/hr} brings some unique information that changes the behaviour of the PNS, similarly as it does for PS in the third row. We attribute this feature to the strong autoregressive serial dependency of \texttt{v/hr}.

Interestingly, we observe that when $\lambda = 0$, including \texttt{v/hr} yields similar PN and PS values ($\approx 50\%$) as in Figure \ref{figure:p-f-cf-unif-weights} for $v \approx 2.52$, meaning that the maximisation does not help much for low values of the regularisation parameter $\lambda$. However, when \texttt{v/hr} is excluded, the PS drops to zero whilst the PN is larger than $95\%$ for all values of $\lambda$. This suggests that an extreme in \texttt{v/hr} is necessary to imply extremes in some other marginals in the following six hours although not sufficient, that is, that the probability that an impact extreme would happen when an extreme in \texttt{v/hr} has \emph{not} happened is still relatively high.

\begin{landscape}
\begin{figure}[htp]
  \begin{center}
\includegraphics[width=1.5\textwidth]{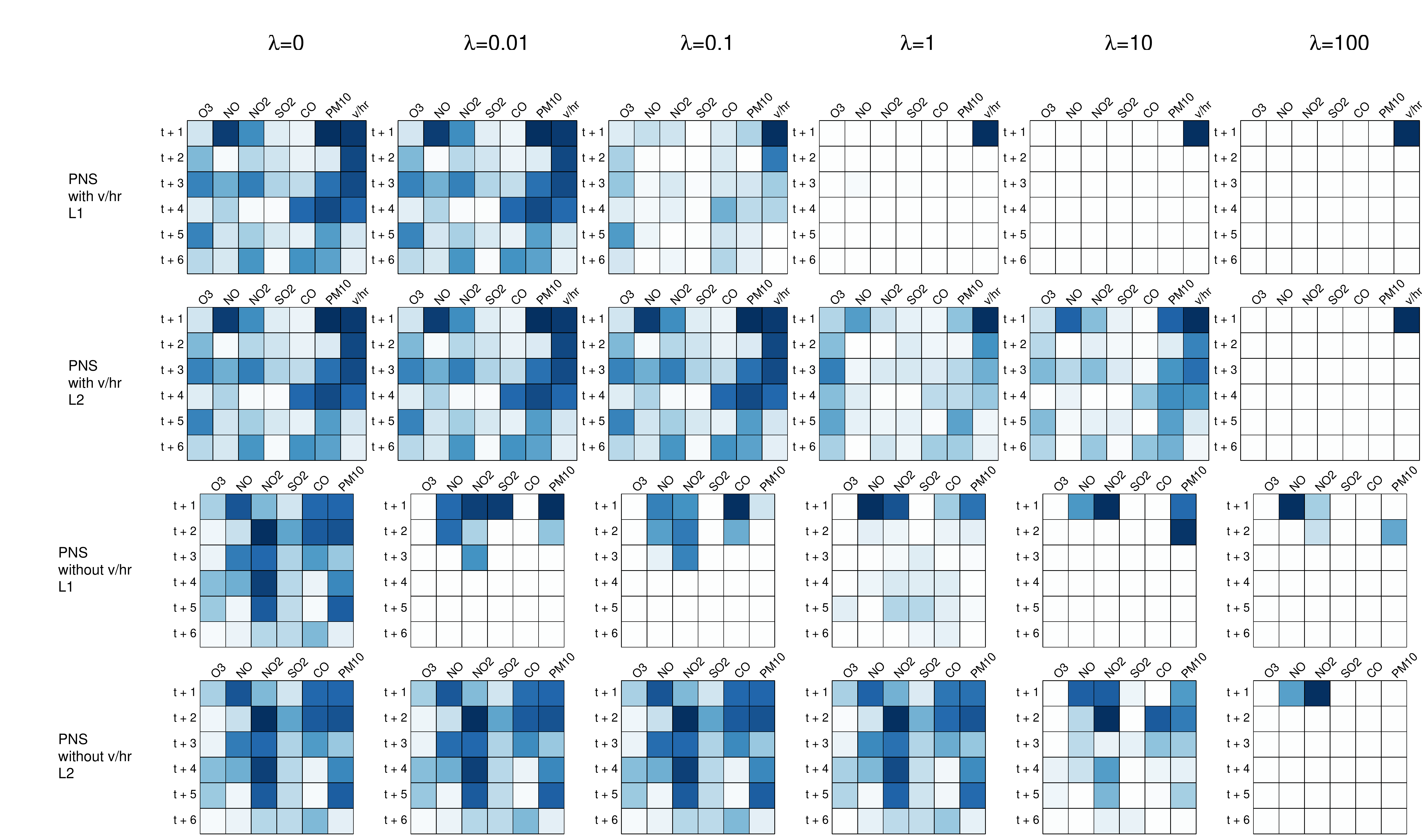}
\caption{PNS maximisation with (top 2 rows) or without (bottom 2 rows)  the cause marginal \texttt{v/hr} in the impact event, as a function of the regularisation parameter $\lambda \in \{0, 0.01, 0.1, 1, 10, 100\}$ (from left to right) and of the norm in the regularisation term with $p=1$ on odd rows and $p=2$ on even rows.}
\label{figure:pns}
\end{center}
\end{figure}
\end{landscape}

\paragraph{Weight matrices} Complementarily, we plot the standardised PNS weight matrices after maximisation in Figure \ref{figure:pns}. A first observation is that the $L^2$ penalty requires a much higher $\lambda$ to produce similar levels of sparsity with only the case $\lambda = 100$ showcasing only a few positive weights as opposed to requiring only that $\lambda \geq 1$ for the $L^1$ penalty cases.

We also notice that, as mentioned above, the presence of \texttt{v/hr} concentrates solely the weights on \texttt{v/hr} at $t+1$ as shown in the two top rows. On the other hand, in the two bottom rows, removing \texttt{v/hr} implies a weight concentration on \texttt{NO}, \texttt{NO2} and a bit on \texttt{PM10} for $t+1$ mostly (for $\lambda = 100$, i.e.\ last column). In addition, for $\lambda =0.1$ with the $L^1$ penalty or $\lambda = 10$ with the $L^2$ penalty, the weights of \texttt{NO}, \texttt{NO2} and \texttt{PM10} are still dominant, but so are the weights of  \texttt{CO} at $t+1$ and $t+2$ as well as $t+2$ and $t+3$, respectively. This means, on the short term, traffic spikes lead to high concentration of nitrate oxides (\texttt{NOX}) and of organic compounds such as \texttt{CO} as described in \cite{crutzen1979role}.

Note that although we presented the presence of the cause marginal (\texttt{v/hr}) and regularisation separately at first, their interplay is key to understand the added value of pruning weaker weights.

Finally, we observe that although both \texttt{CO} and \texttt{SO2} have similarly (cross-)correlations with \texttt{v/hr} (see Fig.\ \ref{figure:correlation}), their causal weights are very different which supports the intuition that probabilities of causation capture dependencies beyond linear relationships.

\subsubsection{Discussion}
\label{section:limitations}
The impact threshold $v \in \R$ is key to determine the levels above which the impact function $h$ is considered to be \emph{extreme}. Choosing a large threshold implies that the number of times the impact event occurs in the dataset will thin out, highlighting the importance of tail probability approximations (Section \ref{section:tail-probabilities-approx}). Also, in general, finding a threshold that is consistent for all weight vectors $\vw$ remains an open problem when the impact event is not obtained by the application at hand (e.g.\ Section 5.1, \cite{bevacqua2017multivariate}). The marginal transformation trick (Section \ref{section:tail-probabilities-approx}) is a potential solution to generate quantities closed under some transformation of interest (e.g.\ weighted sum).

Sparse impact events are obtained as the regularisation term ramps up, with only certain links appearing as significant in this context. However, we see that the strong positive autocorrelation in \texttt{v/hr} leads to high impact weights, not necessarily solely from causal relationships. Although we mention that vine copulas can form a sparse dependence structure, dedicated sparse structures for extremes \citep{engelke2021reviewsparse} offer theoretical guarantees that they retrieve the true tree-based description of true extremal dependencies under the assumptions that those relationships are static in time and unconditioned (outside the counterfactual theory). On the contrary, vine copulas can capture dependencies in a flexible and scalable manner through time and cross-sections but are only equipped with their tail dependence function (may they exist) to quantify their extremal behaviour. Although we presented the theoretical foundations of copulas and their practical implications when it comes to tail modelling (Section \ref{section:multivariate-extremes}), an investigation in the spirit of \cite{engelke2020structure} for vine copulas, linking the information criteria with the extremal structure hence created, remains unexplored to our knowledge. Although this goes beyond the scope of this article, a first step could be to maintain a collection of structures that capture separately factual and counterfactual relationships as mentioned in Section\ref{section:regularisation} and this work is the first step in this direction---at least empirically.

\section{Conclusion}
\subsection{Summary}
In this article, we introduce the Extreme Event Propagation (EEP) framework which deals with the temporal and cross-sectional propagation of a cause extreme event on an impact event. We quantify their relationship through a counterfactual framework \citep{pearl1999probaOfCausation}, equipped with a set of three probabilities of causation where we compare two versions of the world: one where the cause has happened and one where it has not. By doing so, we obtain some information about the "cause" extreme event triggering the impact event at a later time. 

Although the EEP framework is model-agnostic, we explore different properties of multivariate peaks-over-thresholds distributions such as extremal correlations, regularly varying distributions and tail dependence functions that we believe are essential to represent accurately the extremal structure between different time series. We then select a copula-based approach that satisfies most of those properties; where marginals have generalised Pareto-distributed upper tails and are linked together through a stationary and flexible dependence structure, namely using stationary Archimedean vine copulas \citep{nagler2020stationary}. 

We focus on marginal extreme events as the cause of a linear impact extreme event \citep{kiriliouk2019PoTMultivariateGPD} for interpretability purposes. By maximising those said probabilities of causation with respect to the linear projection weight, we obtain which and when are the probable marginals to become extreme. Regularisation in the form of an $L^1$ or $L^2$ penalty is also explored to help extract the most significant causal links. This said, our analysis is applicable to settings where the impact event is fully characterised \citep{bevacqua2017multivariate} or beyond the linear case.

Finally, we apply the EEP framework to an air pollution dataset to study the impact of high road traffic on air pollutant concentration. We retrieve the chemical and physical reactions documented in the literature \citep{crutzen1979role, coria2015airpollution} and observe that regularising the problem helps in generating interpretable (e.g.\ sparse) results about the underlying causal dynamics. 

\subsection{Outlook}

Studying the propagation of extremes through a causal framework poses a number of challenges. Understanding which variables are the most probable to become extreme themselves is currently estimated in two ways: by inspection of the weights or via regularisation. However, devising a hypothesis testing framework or a theoretical understanding of the induced regularisation bias, respectively, to quantify the weight significance remain open problems. By including serial and cross-sectional dependencies at the heart of the EEP framework using stationary vine copulas, we contribute to forming a family of approaches in addition to the current static methods \citep{hannart2018probabilities, kiriliouk2019PoTMultivariateGPD, mhalla2020causalmechanism}. Given the similarity between our modelling framework and recent contributions regarding extreme values through sparse structures \citep{engelke2020structure, engelke2021reviewsparse}, we believe that a formal link between those approaches is bound to be exploited in the near future.

\begin{appendix}

\section{Baseline bivariate copulas or pair-copulas}
\label{section:intro_vine_copulas}

We apply a stationary Markovian vine model with Archimedean or independent bivariate copulas. This allows capturing tail dependence across the marginals themselves. We control the fitting performance and the number of parameters included in this second layer using information criteria such as AIC, BIC or sparse modified BIC for vines (mBICV) \citep{nagler2018model}.

A key limitation of traditional copulas is the incapacity to scale up with the number of variables since those copulas are often parametrised with very few parameters (one or two).

Regular vine construction of copulas is performed using a \emph{constellation of parametric bivariate dependence functions} \citep{grothe2013vine}, see Definition \ref{def:r-vine}, which we call \emph{pair-copulas}. The regular vines pair construction is detailed in Section \ref{section:dissmann-vine-structure}. Since copulas are designed to capture dependency, a common measure of dependence is Kendall's $\tau$ \citep{embrechts2013modelling}.

\begin{table}[htp]
\begin{center}
\begin{tabular}{cccc}
 \specialrule{.1em}{.1em}{.1em}
\multicolumn{1}{c}{\textbf{Name}} &
  \textbf{Copula $C(\vu;\theta)$} &
  \textbf{Generator $\phi(t,\theta)$} &
  \textbf{Parameter $\theta$} \\
  \hline
\textit{Independence} &
  $\prod_{i=1}^d u^{i}$ &
  $e^{-t}$ &
  $\emptyset$ \\
\textit{Clayton} &
  $\left(\sum_{i=1}^d (u^{i})^{-\theta} -(d-1)\right)^{-1/\theta}_+$ &
  $\theta^{-1}(t^{-\theta}-1)$ &
  $\theta \in [-1,\infty) \symbol{92} \{0\}$ \\
\textit{Gumbel} &
  $\exp\left\{-\left\{\sum_{i=1}^d(-\ln u^{i})^{\theta} \right\}^{-1/\theta} \right\}$ &
  $(-\ln t)^{\theta}$ &
  $\theta \in [1,\infty)$ \\
\textit{Frank} &
  $-\theta^{-1}\ln \left(1+\frac{\prod_{i=1}^d\left(e^{-\theta u_i}-1\right)}{(e^{-\theta}-1)^{d-1}}\right)$ &
  $-\ln\frac{e^{-\theta t} - 1}{e^{-\theta} - 1}$ &
  $\theta \in \R\symbol{92}\{0\}$ \\
\textit{Joe} &
  $1-\left[\sum_{i=1}^d (1-u_i)^\theta - \prod_{i=1}^d (1-u_i)^\theta \right]^{-1/\theta}$ &
  $-\ln(1-(1-t)^\theta)$ &
  $\theta \in [1,\infty)$\\
   \specialrule{.1em}{.1em}{.1em}
\end{tabular}
\label{table:archimedean-copulas}
\caption{List of the four Archimedean pair-copulas.}
\end{center}
\end{table}

\begin{remark}
To obtain negative dependence between variables, there are \emph{rotated} copulas to capture better positive and negative dependency: to do so, we use the transformation $u \mapsto 1-u$ on either one or both variables to rotate the scatter plot and fall back to the case where $\tau > 0$.
\end{remark}

\end{appendix}

\begin{funding}
 The first author was supported by EPSRC grant number EP/R014604/1. AV would also like to acknowledge funding by the Simons Foundation.
\end{funding}


\begin{supplement}
We present a short overview of the different notions of time series causality in Appendix \ref{section:time-series-causality}. Core definitions from the counterfactual causal theory of \cite{pearl1999probaOfCausation} that we use the EEP framework are presented in Appendix \ref{section:primer-counterfactual}. Finally, we recall the vine copula definitions and concepts that we leverage in Appendix \ref{section:archimedean-regular-vine-copulas}.

\section{Time series causality}
\label{section:time-series-causality}
We present a short overview of different causality notions for time series. 
\subsection{General context}Understanding causality in the context of time series analysis is often based on the celebrated Granger causality \citep{granger1980causality, granger1988causality} which can be applied and tested in many time series models from autoregressive models \citep[Section 5]{granger1980causality} to copulas \citep{kim2020copulaGranger} but can capture the spurious causal links in the presence of confounding (latent) variables \citep[Section 4]{eichler2013causalMultipleTS}. Granger causality is often compared to Sims causality \citep{sims2972MoneyIncomeCausality, florens1982noncausalitySims}, structural causality \citep{white2010GrangerCausalityDynamic} and intervention causality \citep{eichler2007causal}; the EEP framework is most closely related to the two last items. We refer to  \cite{eichler2013causalMultipleTS} for an insightful discussion and comparison of the four causality frameworks.

\subsection{Interventions and causality}Two usual assumptions regarding time-series causality are that (a) the cause precedes its effects in time; and, (b) manipulations of the cause change the effects \citep[Section 2.c.]{eichler2013causalMultipleTS}, as defined for the celebrated Granger causality \citep{granger1980causality, granger1988causality}. Intervention causality \citep[Section 2.b.]{eichler2013causalMultipleTS} is defined for four different intervention regimes (e.g.\ idle, atomic, conditional \& random) dictating the behaviour of the (non-random) intervention indicator $\zeta^i_t$ which gives information about how the data is generated or observed \citep[Def. 2.1 \& Rem. 2.2]{eichler2007causal}.

In theory, in the EEP framework, we would focus on either the  \emph{random} regime where the conditional distribution of $\XX_{t+1}(k)$ given $\XX_0(t)$ is known; or the \emph{conditional} regime, where $\XX_{t+1}(k)$ is forced to take a value that depends on past observations of $\XX_0(t)$. However, as it often impractical or not feasible to collect data under the interventional regime, we work under the idle regime which coincides with the observational regime, where $\XX$ arises naturally, to model dependencies and leverage this structure to generate data under intervention.

We approximate the conditional intervention regime by leveraging the conditional sampling capabilities of vine copulas, see Section \ref{section:intervention-sampling}. More precisely, we borrow elements from the structural causality framework \citep{white2010GrangerCausalityDynamic} which states that the data is generated according a recursive dynamic structure. For a cause of interest $X_t$, an impact of interest $Y_t$ and a collection of (additional) observed variables $Z_t$ and unobserved ones $U_{X,t}, U_{Y,t}$, where $t \in \N$, we have:
$$X_{t+1} = q_{x,t}\left(X_0(t), Y_0(t), Z_0(t), U_{X=x,t+1}\right),  \quad\text{and}\quad Y_{t+1} = q_{y,t}\left(X_0(t), Y_0(t), Z_0(t), U_{Y=y,t+1}\right),$$
for an unknown function $q_{\cdot, t}$.
This dependence structure resembles that of vine copulas \citep{joe2010tail, bedford2002vines, bedford2001vines} especially when tailored for time series \citep{nagler2020stationary, smith2015dvine, beare2015mvine, brechmann2012truncatedvines}, under the assumption that there are no unobserved variables. 

This collection of links between causality structures and vine copulas explains why they are strong modelling candidates; in addition, we recall the theoretical guarantees and properties of vine copulas for multivariate extreme value modelling in Section \ref{section:multivariate-extremes}. A key component in counterfactual causality modelling is the methodology used to compare different settings, which we discuss in the following section.

 \subsection{An alternative measure of causality}As opposed to using probabilities of causation, causality in the context of time series is usually quantified using the average causal effect (\ACE) \citep[Def.\ 2.1]{eichler2013causalMultipleTS}. Informally, it unveils the impact of the intervention at time $t$ in the sense of an exceedance in expectations between the target variable at time $t+h$  under intervention and without it. It is defined by 
$$\ACE^{(i,j)}_{x}(h) \overset{\Delta}{=} \E(X^{j}_{t+h}\ |\ do(X^{i}_t = x)) - \E(X^{j}_{t+h}),\quad \text{for $x \in \R^d$, $i,j\in I$ and $h \in \N$},$$
 where the expectation is taken under the observed probability measure $\proba$ without interventions as mentioned above. In this setting, \cite{eichler2007causal} discusses causality identifiability: under the conditional or random regimes, if $i,j \in S \subseteq I$ then $S$ identifies the effect of $X^i_t$ on $X^j_{t+h}$ for all $h\in\N$ if an intervention in $X^i_t$ with intervention indicator $\zeta^i_t$ satisfies 
 \begin{equation}
 \label{eq:intervention-indicator-assumptions}
 	\{\XX_0(t-1),\ (X^j_t,\ j\neq i)\} \indep \zeta^i_t, \quad \{\XX_{t+h},\ h \in \N\} \indep \zeta^i_t\ |\ \XX_0(t),\quad \text{and} \quad X^j_{t+h} \indep \zeta^i_t\ |\ \XX_0(t),
 \end{equation}
 for all $h \in \N$, with the (conditional) independence ($\indep$) from \cite{dawid1979conditionalexpectation}, see below. The first assumption ensures that intervening on the past or any other variables at time $t$ which excludes instantaneous causality. The second independence means that the future values are only affected by interventions through past variables; similarly, the third assumption states that the target variables are only affected by interventions through past variables. 
 
 By considering the difference between factual and counterfactual interventions, the ACE can extend the concept of the PNS beyond (and including) binary events. However, the PCs quantify in three different ways the relationship of the cause on the impact events which we believe are better suited to reflect the complex mechanisms involved in the propagation of extremes. That is, they gain value when presented jointly (see Section \ref{section:london-air-pollution-dataset}). Sufficient and necessary causations are thought to be complementary: on page 95, \cite{pearl1999probaOfCausation}, it is explained that
\begin{displayquote}
	$[...]$ necessary causation is a concept tailored to a specific event under consideration, while sufficient causation is based on the general tendency of certain event \emph{types} to produce other event types. Adequate explanations should respect both aspects. If we base explanations solely on generic tendencies (i.e., \emph{sufficient} causation), we lose important specific information. [emphasis in original]
\end{displayquote}

Note that, in this context, the independence $X \indep Y$ holds if and only if $f_{X,Y} = \tilde{f}_{X}\times \tilde{f}_{Y}$ where $\tilde{f}_X, \tilde{f}_{Y}$ are not necessarily the marginal densities $f_X,f_Y$. Similarly, $X \indep Y | Z$ if and only if $f_{X,Y|Z} = \tilde{f}_{X|Z} \times \tilde{f}_{Y|Z}$ where $\tilde{f}_{\cdot | Z}$ are proper distributions. This highlights the fact that any marginal distributions does not depend on the other marginal distributions. Also, $X^j_{t+h} \indep \zeta^i_t$ for some $h \in \N$ means that the distribution of $X^j_{t+h}$ is the same under any of the intervention regimes.

\section{Primer on the counterfactual causal theory}
\label{section:primer-counterfactual}
 \subsection{Causal models, actions and potential response}
 \label{section:causal-models}
Causality requires to separate internal variables of the system under study from external ones as well as the knowledge of how they are related to one another, as formalised in the following definition:
 \begin{definition}{\citep[adapted from Def.\ 1]{pearl1999probaOfCausation}}
 \label{definition:causal-model}
 	A causal model is a triple $M=(U,V,F)$ where
 	\begin{enumerate}[label=(\roman*)]
 		\item $U$ is a set of variables called exogenous that are determined by factors outside the model.
 		\item $V=\{V_1,\dots,V_d\}$ is a set of variables, called endogenous, that are determined by variables in the model, namely, by variables in $U \cup V$.
 		\item $F=\{f_1,\dots,f_d\}$ where each $f_i:U \times (V\symbol{92} V_i) \rightarrow V_i$ gives the value of $V_i$ given the values of all other variables in $U \cup V$ which can be represented by
 		$$v_i = f_i(pa_i, u_i), \quad i \in I,$$
 		where $pa_i$ (resp.\ $u_i$) is any realisation of the unique minimal set of parent variables $\PA_i$ (resp.\ $U_i$) in $V\symbol{92}V_i$ (resp.\ in $U$) that renders $f_i$ nontrivial. 
 	\end{enumerate}
 \end{definition}
 Note that the definition of endogenous variables is recursive and they are fully determined by exogenous variables. A causal model $M$ is commonly associated with a directed graph called the causal graph and denoted $G(M)$, where $V$ are the nodes and the directed edges are from the parent variables in $\PA_i$ towards $V_i$ for any $i \in I$.  
 
 \begin{definition}{\citep[Def.\ 2]{pearl1999probaOfCausation}}
  \label{definition:causal-submodel}
 	Let $M$ be a causal model and $X \subseteq V$, and $x$ be a realisation of $X$. A submodel $M_x$ of $M$ is the causal model $M_x = (U,V,F_x)$, where $F_x \overset{\Delta}{=} \{f_i:V_i \not\in X \} \cup \{X=x\}$. \end{definition}
 A submodel $M_x$ is similar to $M$ where all functions $f_i$ corresponding to variables in $X$ are replaced by constant functions such that $X=x$. Analogously, acting on $M$ by imposing $X=x$ is defined as follows:
 \begin{definition}{\citep[adapted from Def.\ 3]{pearl1999probaOfCausation}}
  \label{definition:causal-action}
 	Let $M=(U,V,F)$ be a causal model, $X \subseteq V$ and $x$ be a realisation of $X$. We define the action $do(X=x)$ as the minimal change in $M$ required to make $X=x$ hold true under any $u \in U$.  The effect of action $do(X=x)$ on $M$ is given by the submodel $M_x$.
 \end{definition}

\begin{figure}[htp]
\centering
\begin{tikzpicture}[node distance={17mm}, thick, main/.style = {draw, circle}, minimum size=.4cm] 
\begin{scope}
\node[main] (1) {$V_A$}; 
\node[main] (2) [above right of=1] {$V_B$}; 
\node[main] (3) [below right of=1] {$V_C$}; 
\node[main] (4) [above right of=3] {$V_D$}; 
\node[main,rectangle,draw] (5) [left of=1] {$U_A$}; 
\node[main,rectangle,draw] (6) [left of=2] {$U_B$}; 
\node[main,rectangle,draw] (7) [left of=3] {$U_C$}; 
\node[main,rectangle,draw] (8) [right of=4] {$U_D$}; 
\Edge[Direct](1)(2);
\Edge[Direct](1)(3);
\Edge[Direct](2)(4);
\Edge[Direct](3)(4);
\Edge[Direct, style=dashed, distance=4.7](5)(1);
\Edge[Direct, style=dashed, distance=4.7](6)(2);
\Edge[Direct, style=dashed, distance=4.7](7)(3);
\Edge[Direct, style=dashed, distance=4.7](8)(4);
\end{scope} 

\begin{scope}[xshift=8.5cm]
\node[main] (1) {$V_A$}; 
\node[main,rectangle,draw] (2) [above right of=1] {$v_B$}; 
\node[main] (3) [below right of=1] {$V_C$}; 
\node[main] (4) [above right of=3] {$V_D$}; 
\node[main,rectangle,draw] (5) [left of=1] {$U_A$}; 
\node[main,rectangle,draw] (7) [left of=3] {$U_C$}; 
\node[main,rectangle,draw] (8) [right of=4] {$U_D$}; 
\Edge[Direct](1)(3);
\Edge[Direct](2)(4);
\Edge[Direct](3)(4);
\Edge[Direct, style=dashed, distance=4.7](5)(1);
\Edge[Direct, style=dashed, distance=4.7](7)(3);
\Edge[Direct, style=dashed, distance=4.7](8)(4);
\end{scope} 
\end{tikzpicture} 
\caption{Intervention $do(V_B=v_B)$ on $U=\{U_A,U_B,U_C,U_D\}$ and $V=\{V_A,V_B,V_C,V_D\}$.}
\end{figure}
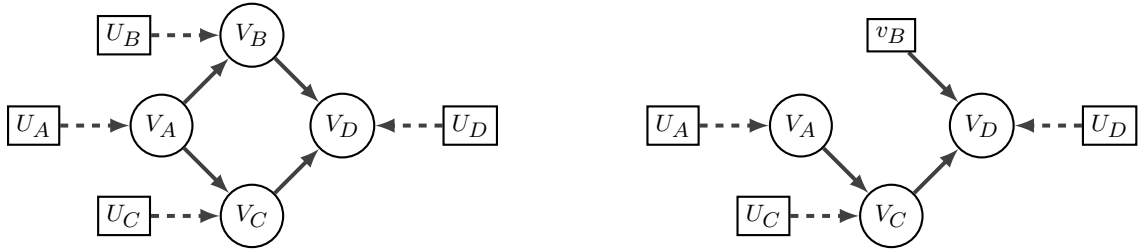

 Given the effect of action $do(X=x)$, we recall the definition of the potential response
 \begin{definition}{\citep[Def.\ 4 \& 5]{pearl1999probaOfCausation}}
   \label{definition:potential-response}
 	Let $Y\in V$, and let $X \subseteq V$. The potential response of $Y$ in unit $u$ to action $do(X=x)$, denoted $Y_x(u)$, is the solution for $Y$ of the set of equations $F_x$. The counterfactual sentence "the value that $Y$ would have obtained, had $X$ been $x$." is interpreted as denoting the potential response $Y_x(u)$.
 \end{definition}

 Furthermore, we consider the extension to probabilistic causal models defined as follows
 \begin{definition}{\citep[Def.\ 6]{pearl1999probaOfCausation}}
   \label{definition:probabilistic-causal-model}
 	A probabilistic causal model is a pair $(M, \proba(u))$, where $M$ is a causal model and $\proba(u)$ is a probability function defined on the domain of $U$.
 \end{definition}
 
 From Def.\ \ref{definition:causal-model}, all endogenous variables $V$ are functions of exogenous variables $U$. In that sense, for any $Y \subseteq V$, we write 
 $$\proba(Y=y) = \sum_{u:\ Y(u) = y} P(u), \quad \text{and} \quad \proba(Y_x = y) = \sum_{u:\ Y_x(u) = y}\proba(u),$$
 and similarly for all probabilities involving variables in $V$.
 
 \begin{notation}
We write $\proba\left(Y \ |\ do(X = x)\right)$ for $\proba\left(Y_x\right)$ to emphasise the action applied on $Y$. 
 \end{notation}
 
 \subsection{Probabilities of causation}
 
 Given the definition of causal models and counterfactuals, we define three probabilities of causation ($\PC$) \citep{pearl1999probaOfCausation, hannart2016causal, hannart2018probabilities} which outline different relationships between some action on a causal model and its corresponding response.
 
 \begin{definition}{\citep[Def.\ 7, 8 \& 9]{pearl1999probaOfCausation}}
 \label{definition:causal-probabilities}
 	Let $X$ and $Y$ be two binary variables in a causal model $M$. The probabilities of necessary, sufficient and necessary and sufficient causation are defined as the expressions
 	\begin{align*}
 		\PN &\overset{\Delta}{=} \proba\left(Y = 0\ |\ do(X=0),\ X=1,\ Y=1\right), \quad &&\textit{(necessary),}\\
 		\PS  &\overset{\Delta}{=} \proba\left(Y = 1\ |\ do(X=1),\ X=0,\ Y=0\right), \quad &&\textit{(sufficient),}\\
 		\PNS &\overset{\Delta}{=} \proba\left(\left\{ Y = 0\ |\ do(X=0)\right\} \cap \left\{Y = 1\ |\ do(X=1)\right\}\right), \quad &&\textit{(necessary and sufficient).}	
 	\end{align*}
 \end{definition}
We note that the PC quantities are linked through the relationship \citep[Lemma 1]{pearl1999probaOfCausation}
  $$\PNS = \proba(X=1,Y=1) \cdot \PN + P(X=0, Y=0) \cdot \PS.$$
Necessary causation ($\PN$) is defined as the likelihood that $Y$ would be zero had $X$ been $0$, given that, in reality both $Y$ and $X$ are actually $1$. Sufficient causation $(\PS)$ is the opposite: it is the likelihood that $Y$ would be $1$ had $X$ been $1$, when both $X$ and $Y$ are actually equal to $0$. The probability of necessary and sufficient causation ($\PNS$) sits in between as the likelihood that $Y$ is equal to $1$ had $X$ been $1$ and that $Y$ is $0$ had $X$ been equal to $0$. We translate those probabilities into the EEP framework in Section \ref{section:causal-probabilities-eep}.
Those probabilities can be defined on a causal model $M$ which can or cannot be identifiable in the following sense:
\begin{definition}{\citep[Def.\ 12]{pearl1999probaOfCausation}}
\label{definition:identifiability}
	Let $Q(M)$ be any quantity defined on a probabilistic causal model $(M, \proba)$. $Q$ is identifiable in a class $\gM$ of causal models if and only if any two models $(M_1, \proba_{M_1}),(M_2,\proba_{M_2})\in \gM$ that satisfy $\proba_{M_1} = \proba_{M_2}$ also satisfy $Q(M_1)=Q(M_2)$. In other words, $Q$ is identifiable if it can be determined uniquely from the probability distribution $\proba$ of the endogenous variables $V$.
\end{definition}

\begin{notation}
	Let $x$ stand for $\{X=1\}$ and, $x^c$ for its complement $\{X=0\}$. 
\end{notation}
To ensure the identifiability of all three $\PC$, it is usual to assume the \emph{exogeneity} and \emph{monotonity} of $X$ with respect to $Y$ \citep[Section 3.3]{pearl1999probaOfCausation} which we recall below:

  \begin{definition}{\citep[Def.\ 13]{pearl1999probaOfCausation}}
  \label{definition:exogeneity}
  	A variable $X$ is said to be exogenous relative to $Y$ in a causal model $M$ if and only if (iff) $\proba(Y_x = 1,\ Y_{x^c} = 1\ |\ x) = \proba(Y_x=1,\ Y_{x^c}=0)$, i.e.\ iff the potential response of $Y$ to the action $x$ or $x^c$ is independent of the actual value of $X$.
  \end{definition}
 A key property of exogeneity is the identification of $\proba(Y|do(X))$ to the corresponding conditional probability $\proba(Y|X)$ which allows for computation using empirical data, see Section \ref{section:time-series-causality}.

\begin{definition}{\citep[Def.\ 14]{pearl1999probaOfCausation}}
	\label{definition:monotonocity}
	A variable $Y$ is said to be monotonic relative to variable $X$ in a causal model $M$ iff the junction $Y_x(u)$ is monotonic in $x$ for all realisation $u$ of $U$. Equivalently, $Y$ is monotonic relative to $X$ iff $\proba(Y_{x}=0, \ Y_{x^c} = 1) = 0$.
\end{definition}
That is, if $Y$ is monotonic relative to $X$, then $Y$ can only move in the same direction as $X$: a condition change from $x^c$ to $x$ (i.e.\ $X$ grows from 0 to 1) will not change $Y$ from $y$ to $y^c$,i.e.\ decreases from $1$ to $0$, irrespective of the exogenous variables. Under both exogeneity and monotonity conditions, all three $\PC$ are identifiable \citep[Th.\ 3]{pearl1999probaOfCausation}.

\section{Archimedean Stationary Vine Copulas}
\label{section:archimedean-regular-vine-copulas}

We call Archimedean stationary vine copulas (ASVC) any stationary vine copulas constructed using either Archimedean or independent bivariate copulas which act as building blocks and propagate dependencies through the vine trees.

\subsection{Definitions}
We define a tree $T=(N,E)$ as an acyclic graph, where $N$ is its set of nodes and $E$ is its set of edges (unordered pairs of nodes), see Def.\ 4, \cite{bedford2001vines}. 
\subsubsection{Vine copulas}
We recall the definitions of vines and regular vine copulas (or R-Vine):
\begin{definition}{\citep[Def.\ 8]{bedford2001vines}}
\label{def:r-vine}
A vine $\gV$ on $d$ elements is an ordered sequence of trees $\mathcal{T} \overset{\Delta}{=} (T_1, \dots , T_{m})$ with $T_l \overset{\Delta}{=} (N_l,\ E_l),$ for $l \in \{1,\dots,m\}$ such that:
\begin{enumerate}[label=(\roman*)]
    \item $N_1 \overset{\Delta}{=} \{1,\dots,d\}$ i.e.\ the first tree has nodes 1,\dots,d;
    \item for $l \geq 2$, $T_l$ is a tree with nodes $N_l \subset N_1 \cup E_1 \cup \dots \cup E_{l-1}$.
\end{enumerate}
A vine $\gV$ is a regular vine on $d$ elements if

\begin{enumerate}[label=(\roman*)]
\item it is made exactly $d$ trees, i.e.\ $m$ = $d$;
\item $T_l$ is a connected tree with a node set equal to the edge set of the previous tree, i.e.\ $N_l \overset{\Delta}{=} E_{l-1}$ with $\card(N_l) = d-i+1$ for $d \in I$, where $\card(N_l)$ is the cardinality of the set $N_l$;
\item the proximity condition holds: for $l \in \{1,\dots,d-1\}$, if two nodes from $N_{l+1}$ are connected in $T_{l+1}$, the corresponding edges in $T_l$ have exactly one common node.
\end{enumerate}
\end{definition}

In theory, each pair copula of the vine is fitted conditional on the uniform bivariate distribution of the previous level. However, it is usual in the literature to make the assumption that we only fit given the uniform univariate marginals themselves to make the computation more tractable.

If the trees $T_i$ correspond to a path (i.e.\ where each node has either one or two neighbours), the vine is called a \emph{D-vine}. On the other hand, if the trees are stars (i.e.\ all but one nodes have the same unique neighbour), it is a \emph{C-vine}.

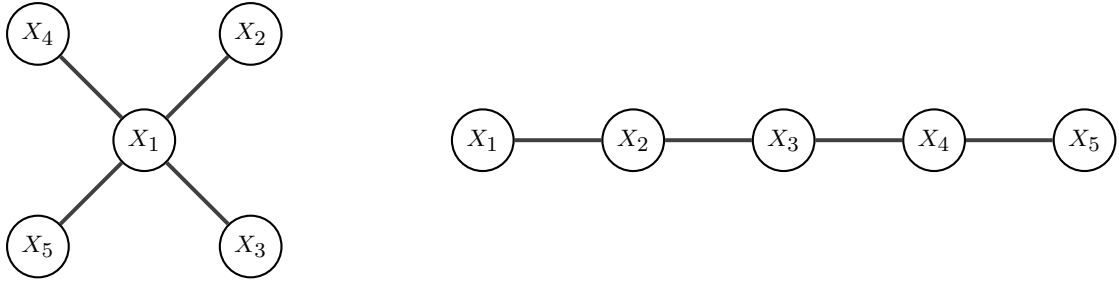
\begin{figure}
\centering
\begin{tikzpicture}[node distance={20mm}, thick, main/.style = {draw, circle}, minimum size=.1cm] 
\begin{scope}
\node[main] (1) {$X_1$}; 
\node[main] (2) [above right of=1] {$X_2$}; 
\node[main] (3) [below right of=1] {$X_3$}; 
\node[main] (4) [above left of=1] {$X_4$}; 
\node[main] (5) [below left of=1] {$X_5$}; 
\Edge[](1)(2);
\Edge[](1)(3);
\Edge[](1)(4);
\Edge[](1)(5);
\end{scope} 

\begin{scope}[xshift=4.5cm]
\node[main] (1) {$X_1$}; 
\node[main] (2) [right of=1] {$X_2$}; 
\node[main] (3) [right of=2] {$X_3$}; 
\node[main] (4) [right of=3] {$X_4$}; 
\node[main] (5) [right of=4] {$X_5$}; 
\Edge[](1)(2);
\Edge[](2)(3);
\Edge[](3)(4);
\Edge[](4)(5);
\end{scope} 
\end{tikzpicture} 
\caption{The first tree ($T_1$) of a (left) D-vine; (right) C-vine on variables $\{X_1,X_2,X_3,X_4,X_5\}$.}
\end{figure}

\subsubsection{Vine copulas for stationary multivariate time series}

\paragraph{General context}Constraining further those tree structures generates suitable vines to model \emph{stationary} multivariate time series, e.g.\ D-vines of \cite{smith2015dvine}, M-Vines of \cite{beare2015mvine} and S-Vines (or stationary vines) of \cite{nagler2020stationary}. For simplicity, we only consider the latter in this article.

\begin{figure}[htp]
\centering
\begin{tikzpicture}[node distance={20mm}, thick, main/.style = {draw, circle}, minimum size=.1cm] 

\begin{scope}[xshift=4.5cm]
\node[main] (1) {$X^1_1$}; 
\node[main] (2) [right of=1] {$X^2_1$}; 
\node[main] (3) [right of=2] {$X^3_1$}; 
\node[main] (4) [right of=3] {$X^4_1$}; 
\node[main] (5) [right of=4] {$X^5_1$}; 
\Edge[](1)(2);
\Edge[](2)(3);
\Edge[](3)(4);
\Edge[](4)(5);

\node[main] (6) [below of=1]{$X^1_2$}; 
\node[main] (7) [right of=6] {$X^2_2$}; 
\node[main] (8) [right of=7] {$X^3_2$}; 
\node[main] (9) [right of=8] {$X^4_2$}; 
\node[main] (10) [right of=9] {$X^5_2$}; 
\Edge[](6)(7);
\Edge[](7)(8);
\Edge[](8)(9);
\Edge[](9)(10);

\Edge[](4)(9);

\node[main] (11) [below of=6]{$X^1_3$}; 
\node[main] (12) [right of=11] {$X^2_3$}; 
\node[main] (13) [right of=12] {$X^3_3$}; 
\node[main] (14) [right of=13] {$X^4_3$}; 
\node[main] (15) [right of=14] {$X^5_3$}; 
\Edge[](11)(12);
\Edge[](12)(13);
\Edge[](13)(14);
\Edge[](14)(15);

\Edge[](9)(14);

\end{scope} 
\end{tikzpicture} 
\caption{The first tree ($T_1$) of a Markov stationary vine with order $3$ on $\{X^1_t,X^2_t,X^3_t,X^4_t,X^5_t\}$ with cross-sections modelled using a D-vine.}
\label{figure:svine-example}
\end{figure}

\begin{figure}[htp]
  \begin{center}
\includegraphics[width=1.0\textwidth]{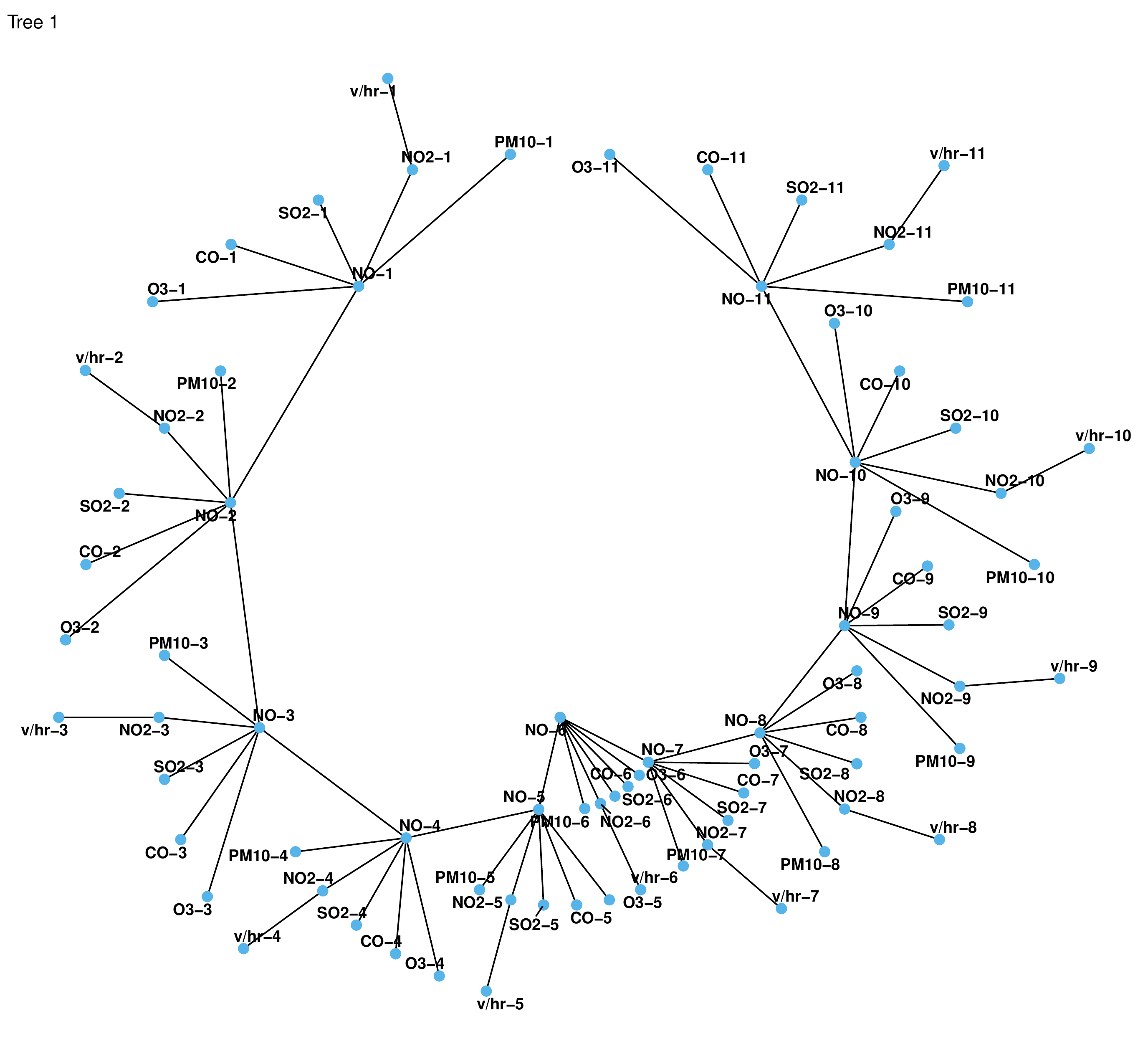}
\caption{First tree of the stationary vine copula fitted on the air pollution dataset where the name of each variable is followed by their time horizon: \text{O3-3} is the third value of Ozone in each 11-long data block.}
\label{figure:tree}
\end{center}
\end{figure}

All those models leverage a single vine that captures the cross-sectional dependence structure of $\XX_t$ at all time points. Then, the first trees of those cross-sectional vines at time $t$ and $t+1$ are linked through a collection of edges with a vertex from the structure at time $t$ and one at time $t+1$, such that those edges are time-invariant. We consider stationary vines in this article where the cross-sectional structure is an arbitrary R-vine and the vine copula model is \emph{translation invariant} \citep[Def.\ 4]{nagler2020stationary}. For instance, in Figure \ref{figure:tree}, we observe that along the backbone $\text{NO-1},\dots,\text{NO-11}$, the cross-sectional vine structure remains the same similar to the D-vine in Fig.\ \ref{figure:svine-example}.

\paragraph{Definitions} We introduce the main definitions necessary for the definition of stationary vine copulas. The interested reader might find it useful to consult Section 2, \cite{nagler2020stationary} for a comprehensive description of stationary vine copulas.

Recall that a vine copula model associates each edge of an regular vine with a bivariate copula. We write $(\gV, \gC(\gV))$ the vine structure $\gV = \{(N_l, E_l),\ l\in \{1,\dots,m\}\}$ and associated collection of bivariate copulas $\gC(\gV)  \overset{\Delta}{=} \{c_e: e \in E_k, k \in\{1,\dots,d-1\}\}$ where $d$ is the number of variables.

\begin{definition}{\citep[Def.\ 4]{nagler2020stationary}} A vine copula model $(\gV, \gC(\gV))$ on the set $N_1 = \{1,\dots,N\} \times \{1,\dots,d\}$ is called translation invariant if it holds that
$$c_{a_e, b_e|D_e} = c_{a_{e'}, b_{e'}|D_{e'}},$$
for all edges $e,e'\in \in_{k=1}^{nd-1}E_k$ for which there is a $\tau \in \ZZ$ such that
$$a_e = a_{e'} + (\tau,0), \quad b_e = b_{e'} + (\tau,0), \quad D_e = D_{e'} + (\tau,0),$$
where the last equality is short for $D_e = \{v+(\tau,0): v \in D_{e'}\}$.
\end{definition}
This definition is the equivalent of the strong stationary condition for vine copulas. The \emph{restriction} and \emph{translation} of vines relate to the translation invariance and we recall their definition:

\begin{definition}{\citep[Def.\ 5]{nagler2020stationary}} Let $(\gV, \gC(\gV))$ be a vine on $\{1,\dots,N\} \times \{1,\dots,d\}$ and let $N_t' = \{t,\dots,t+m\} \times \{1,\dots,d\}$ for some $t,m$ such that $1\leq t \leq N$ and $0\leq m \leq N-t$. For all $k \geq 1$, define $E_k' \overset{\Delta}{=} E_k \cap {{N_k'} \choose {2}}$ and $V_k' \overset{\Delta}{=} E_k'$. Then the sequence of graphs $\gV_{t,t+m} \overset{\Delta}{=} \{(N_k', E_k'): k\in\{1,\dots,(m+1)d-1\}\}$ is called restriction of $\gV$ on the time points $t,\dots,t+m$.
\end{definition}

\begin{definition}{\citep[Def.\ 6]{nagler2020stationary}} Let $m \geq 0$, $\gV_{t,t+m} = \{(N_k', E_k'): k\in\{1,\dots,(m+1)d-1\}\}$ to be a vine on $\{t,\dots,t+\tau\}\times \{1,\dots,d\}$ and $\gV_{s,s+m}$ a vine on $\{s,\dots,s+\tau\}\times \{1,\dots,d\}$. We say that $\gV_t$ is a translation of $\gV_s$ (denoted by $\gV_t \sim \mV_s$) if for all $k\in \{1,\dots,d-1\}$ and edges $e \in E_{t,k}$, there is an edge $e'\in E_{s,k}$ such that $e=e'+(t-s,0)$ (and vice versa).
\end{definition}

Then, the stationary vines are defined through a characterisation given by the following theorem
\begin{theorem}{\citep[Th.\ 1]{nagler2020stationary}}
\label{th:svines}
	Let $\gV$ be a vine on the set $\{1,\dots,n\}\times \{1,\dots,d\}$. Then, the following statements are equivalent:
 	\begin{enumerate}[label=(\roman*)]
 		\item The vine copula model $(\gV, \gC(\gV))$ is stationary for all translation invariant choices of $\gC(\gV))$.
 		\item  \label{def:stationary-vines} There are vines $\gV^{(m)}$, $m=1,\dots,n-1$, defined on $\{1,\dots,m+1\}\times\{1,\dots,d\}$, such that for all $m=0,\dots,n-1$, and $1\leq t \leq n-m$, we have
 		$$\gV_{t,t+m} \sim \gV^{(m)}.$$
 	\end{enumerate}
\end{theorem}
Stationary vine copulas are all vine copulas that satisfy Th.\ \ref{th:svines}, \ref{def:stationary-vines} as it describes the notion of strong stationarity for graphs:
\begin{definition}
	A vine $\gV$ on the set $\{1,\dots,N\}\times \{1,\dots,d\}$ is called stationarity if it statisfies condition \ref{def:stationary-vines}, Th.\ \ref{th:svines}.
\end{definition}

\subsection{Existence and uniqueness of vine copulas}
A vine copula can be seen as a \emph{proper} hierarchical copula as it boils down to the existence and uniqueness of a bivariate copula \citep{sklar1959fonctions} as explained in Section 4.2, \cite{czado2010pair}. The vine copulas made of given bivariate copula families provide a multi-parameter augmented coverage of a subspace of $d$-dimensional copulas of the given families. As a special case of Archimedean copulas, we quote Section 4.2.2, \cite{czado2010pair}:
\begin{displayquote}
 This construction of multivariate distributions and copulas is very general and
flexible since we can use any bivariate copula as a building block in the pair-copula construction model.
In contrast to the extended multivariate Archimedean copulas no restriction to the
Archimedean pair-copulas or further parameter restrictions are necessary.
\end{displayquote}

Vines are fitted using uniform margins and we take a semiparametric transformation approach using an empirical distribution function below the extreme thresholds $(\mu^{i}, \ i \in I)$ and the asymptotic GPD above this said threshold \citep{coles1991modellingExtremeEvents}. Another possibility is to use the empirical distribution function throughout; see \cite{shih1995semiparametricCopulaTransfo} for a comparison study.
\end{supplement}


\bibliographystyle{imsart-nameyear} 
\bibliography{bib_file_model}       

\end{document}